\documentclass[10pt,letterpaper,journal]{IEEEtran}
\usepackage[T1]{fontenc}
\usepackage[latin9]{inputenc}

\usepackage{amsmath}
\usepackage{graphicx}
\usepackage{graphics}
\usepackage{epsfig,color}
\usepackage{amssymb}
\usepackage{cite}
\usepackage{bm}
\usepackage{enumerate}
\usepackage{setspace}
\usepackage{empheq}

\makeatletter
\newtheorem{remrk}{Remark}

\newtheorem{lem}{Lemma} 
 
\newtheorem{theorem}{Theorem}
\newtheorem{ass}{Assumptions}

\usepackage{color}		
\usepackage{epsfig}
\allowdisplaybreaks

\def\wo{{\bm{h}}}

\def\uo{{\bm{a}}}

\def\wot{{\underline{\bm{h}}}}

\def\N{{\mathcal{N}}}

\def\R{{\mathbb{R}}}
\def\Z{{\mathbb{Z}}}
\newcommand{\Ss}{\widehat{\mathcal{S}}}

\newcommand{\supp}{\mathrm{supp}}
\newcommand{\So}{{\mathcal{S}}}

\title{Greedy Sparsity--Promoting Algorithms  for Distributed Learning}
\author{Symeon Chouvardas,   Gerasimos Mileounis,   Nicholas Kalouptsidis,   Sergios Theodoridis,  
\thanks{S.~Chouvardas, N.~Kalouptsidis and S.~Theodoridis are with the Department of Informatics and Telecommunications, University of Athens, Panepistimioupolis,
Ilissia, 157 84 Athens, Greece (e-mail: {schouv,kalou,stheodor}@di.uoa.gr).}
\thanks{G.~Mileounis is with Accenture Digital, Analytics (e-mail: gmil@di.uoa.gr)}}

\begin{document}
\maketitle
\begin{abstract}
This paper focuses on the development of novel greedy techniques for distributed learning under sparsity constraints. Greedy techniques have widely been used in centralized systems due to their low computational requirements and at the same time their relatively good performance in estimating sparse parameter vectors/signals.  The paper reports two new algorithms in the context of sparsity--aware learning.
 In both cases, the goal is  first to identify the support set of the unknown signal  and then to {estimate} the non--zero values    restricted  {to} the active support set. First, an iterative greedy multi--step
 procedure is developed, based on a neighborhood cooperation strategy, using batch processing on the observed data. Next, an extension of the algorithm to the online setting, based on the diffusion LMS rationale for adaptivity, is derived. Theoretical analysis of the algorithms is provided, \textcolor{black}{where it is shown that the batch algorithm converges to the unknown vector if 
 a Restricted Isometry Property (RIP)  holds. Moreover, the online version converges 
  in the mean to the solution vector under some general assumptions}. 
 Finally, the proposed schemes are tested against recently developed sparsity--promoting algorithms and their enhanced performance is verified via simulation examples.
\end{abstract}
\begin{keywords}
Distributed systems, Compressed sensing,
System identification, Greedy algorithms, Adaptive filters.
\end{keywords}
\section{Introduction}
Many real--life signals and systems
 can be described by parsimonious models consisting of
 few non--zero coefficients. 
 Typical examples  include: 
 image and video signals, acoustic signals, echo cancellation, wireless multipath channels,  
 High Definition TV, just to name but a few, \textit{e.g.,} \cite{BrDoEl09,ThKoSl13,M10,babadi2010sparls,CMG14,Theo15}.
This feature is of particular importance under the big data paradigm and the three associated dimensions (Volume, Velocity, Variety)  \cite{Lin13}, \textcolor{black}{ where the resulting data set can not be processed as is and exploitation of significant variables becomes crucial}. The goal of this paper is to develop sparsity--promoting algorithms for the estimation
of sparse signals and systems in the context
of distributed environments. It is anticipated that  sparsity--aware learning will constitute a major pillar for big data analytics, which currently seek for decentralized processing and fast execution times \cite{Lin13,boyd11}.

Two of the main paths for developing schemes  for sparsity--aware learning, are a) regularization of the cost via the $\ell_1$ norm of the parameter vector (Basis pursuit), \textit{e.g.,} \cite{TW10,ThKoSl13}, and b) via the use of   greedy algorithms (or Matching Pursuit)  \cite{Tr04,DTDS09,NeedTro09,DaMi09,NV09,HM11,PEE12,Tem11}. In the greedy schemes, the support set, in which the non--zero coefficients lie is first identified and then the respective coefficients are estimated in a step--wise manner. Each one of the above algorithmic families poses their own advantages and disadvantages. The $\ell_1$--regularization methods provide well established recovery conditions, \textcolor{black}{albeit at the expense of} 
longer execution times and offline fine tuning. On the contrary, greedy algorithms perform at lower computational demands; moreover, they still enjoy theoretical justifications for their performance, 
under some  assumptions.

With only few exceptions, \textit{e.g.,} \cite{MaBaGi10,MoXaAqPu10,PaElKe13,LoSa12,ChoSlaKoT12,liu2014distributed}, sparsity promoting algorithms assume that the training data, through which the unknown target vector is estimated, are centrally available. That is, there exist a central processing unit (or Fusion Center, FC), which has access to all training data, and performs all essential computations. Nevertheless, in many applications the need for decentralized/distributed processing rises. Typical examples of such applications are those involving wireless sensor networks (WSNs), which are deployed over a geographical region or software agents monitoring software operations over a network to diagnose and prevent security threats, and the nodes are tasked to collaboratively estimate an unknown (but common) sparse vector. The existence of a fusion center, which collects all training data and then performs the required computations, may be prohibited due to geographical constraints and/or energy and bandwidth limitations. Furthermore, in many cases, \textit{e.g.,} medical applications, privacy has to be preserved which means that the nodes are not allowed to exchange their learning data but only their learning results \cite{CKMJXM02}. \textcolor{black}{Hence}, the development of distributed algorithms is of significant importance.

 This paper is concerned  with the task of sparsity--aware learning, via 
 the greedy avenue, by considering both \textcolor{black}{batch} as well as online schemes. In the former, it is assumed that each node 	
 possesses a \textit{finite} number of input/output measurements, 
 which are related via  the unknown sparse vector. The support set
  is computed in a collaborative way; that is, 
  the nodes  exchange their measurements and exploit them
  properly so as to identify the positions of the non--zero coefficients. 
  In the sequel, the produced estimates, 
  which are restricted \textcolor{black}{to} the
  support set, are fused  according to a predefined rule. 
  The online scenario deals with an \textit{infinite}
  number of observations,  received sequentially at 
  each node. In general, adaptive/online algorithms update the estimates of the
  unknown vector dynamically, in contrast to their batch counterparts,
  where for every new pair of data,   estimation of the unknown vector
  is repeated from scratch. Such techniques can handle cases where the
  unknown signal and its corresponding support set is time--varying  and also cases where the number of available data is very large and the computational resources are not enough
   for such  handling. In the greedy--based adaptive algorithm, which
  is proposed here, the nodes cooperate in order to estimate their support sets,
  and the diffusion rationale, \textit{e.g.,} \cite{sayedarxiv},
  is adopted for the unknown vector estimation.

The paper is organized as follows. In Section II, the problem formulation is described.
 In Section III the
distributed greedy learning problem is discussed and the batch learning algorithm is provided.
Section IV presents  the adaptive distributed learning problem, together with the proposed adaptive scheme
and in Section V the theoretical properties of the presented algorithms are discussed. 
Finally, in Section VI,  the performance of the proposed algorithmic schemes is validated. Details of the derivations are given in the appendices. 

\textit{Notation:}  The set of all  non--negative integers and the set of all real numbers will be denoted by  $\Z$ and $\R$ respectively. Vectors and matrices will be denoted by boldface letters and uppercase boldface letters respectively.
The Euclidean norm will be denoted by $\Vert \cdot \Vert_{\ell_2}$ and the $\ell_1$ norm by $\Vert \cdot \Vert_{\ell_1}$. Given a vector $\bm{h}\in\R^m$, the operator
$\supp_s(\bm{h})$ returns the subset of the $s$ largest in magnitude coefficients. 
 Moreover, $\mathbb{E}[\cdot]$ stands for the expectation operator.  Finally,  $\vert\mathcal{S}\vert$ denotes the cardinality of the set $\mathcal{S}$.


\section{Problem Formulation}
\label{sec:probform}
\begin{figure*}[!t]

  \centering
\includegraphics[scale=0.6]{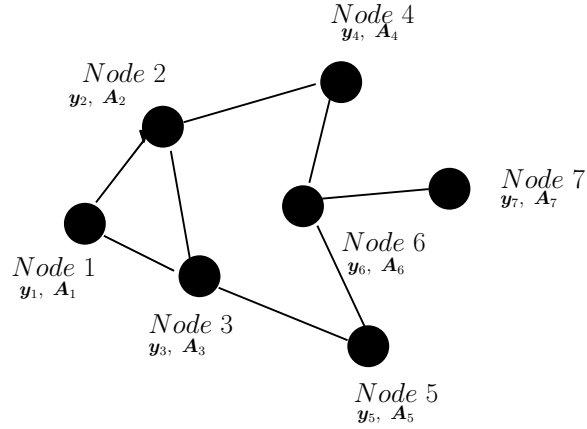}
  
%
%
%
\caption{ An ad--hoc network.\hspace{1.0 pt}}\medskip
\label{fig:diff}
\end{figure*}

Our task is to estimate an unknown  sparse parameter vector, $\boldsymbol{h}_*\in \R^m$,
exploiting   measurements, collected at the $N$ nodes of an ad--hoc network, under a preselected cooperative protocol. \textcolor{black}{The greedy philosophy will be employed.} Such a network is 
illustrated in Fig. \ref{fig:diff}.
We denote the node set by $\N = \lbrace 1,\ldots,N\rbrace$, and we assume that each node is able to exchange information
with a subset of $\N$, namely $\N_k\subseteq\N, k = 1, \ldots ,N$. This
set is also known as the \textit{neighborhood} of $k$. The input--output relation adheres
to the following linear model: 
\begin{equation}
\boldsymbol{y}_{k}=\boldsymbol{A}_{k}\boldsymbol{h}_*+ \boldsymbol{\eta}_k, \ \forall k\in\N,
\label{linsys}
\end{equation}
where $\boldsymbol{A}_{k}$ is an $l\times m$ sensing matrix, with $l < m$, $\boldsymbol{y}_{k}\in \R^l$ and $ \boldsymbol{\eta}_k\in \R^l$ 
is  {an additive white} noise process. The vector to be estimated is assumed to be at most $s$--sparse, \textit{i.e.,} $\Vert\boldsymbol{h}_*\Vert_{\ell_0}\leq s \ll m$,
where  $\Vert\cdot\Vert_{\ell_0}$ denotes the $\ell_0$ (pseudo) norm. 
 \subsection{Sparsity--Aware Distributed Learning: The DLasso Algorithm}
\textcolor{black}{A version of the celebrated Lasso algorithm, in the context of 
distributed learning, is the so--called Distributed Lasso (DLasso) algorithm,
proposed in \cite{MaBaGi10}.
This algorithm  solves the following problem:} 
 	 \begin{equation}
	 \hat{\boldsymbol{h}}=\arg\min_{\Vert \boldsymbol{h} \Vert_{\ell_1}\leq \delta} \sum_{k\in\N}\Vert \boldsymbol{y}_k-\boldsymbol{A}_k\boldsymbol{h} \Vert_{\ell_2}^2.
	 \label{dlasso}
	 \end{equation} 
This optimization can be reformulated  in a fully decentralized way,
where each node exploits local information, 
as well as information received by the neighborhood.
The decentralized optimization can be efficiently solved
 by adopting
the Alternating Direction--Method of Multipliers (ADMM), \cite{BeTs99}. 
It has been shown in \cite{MaBaGi10},
that the solution \textcolor{black}{of the decentralized task} where each node has
\textcolor{black}{access to locally obtained observations
as well as estimates collected by its neighborhood},
 coincides with the solution obtained by solving  
(\ref{dlasso}), which requires global information. 
It should be pointed out that  each node lacks global network information; however,  if an 
appropriate cooperation protocol is adopted, then 
the global solution can be obtained. The adopted cooperation protocol, which
 is based on the ADMM and can assure such a behaviour, 
 is known as \textit{consensus} protocol
  and constitutes the beginning for several distributed learning algorithms, \textit{e.g.,}   
 the distributed Support Vector Machines,  \cite{FoCaGi10}, etc.

\section{Distributed greedy learning}\label{sec:DiGreeLe}
 Greedy algorithms,    \textit{e.g.,} \cite{Tr04,DTDS09,NeedTro09,DaMi09,NV09,HM11,PEE12,Theo15}, 
 offer an alternative sparsity-promoting route to $\ell_1$--regularization; 
 our goal is to extend  the philosophy behind such algorithms in the framework
  of distributed learning.
 
 Under the centralized scenario,
 greedy techniques iteratively estimate the unknown parameter
 by applying the following two--step approach:
\vspace*{-6pt} 
\begin{itemize}
\item \textbf{Subset Selection}: First, a proxy signal is computed (which carries information regarding the
 support set of the unknown vector) based on  input--output measurements. In the sequel, the $s$ largest in amplitude coefficients of the proxy are computed.   
\item \textbf{Greedy Update}: The estimate of the unknown vector is 
computed, by performing a Least--Squares task restricted \textcolor{black}{to} the identified support set.
\end{itemize}
\vspace*{-6pt}

To keep the discussion simple, let us assume that the signal proxy is computed in an non--cooperative way; that is,
  each node exploits solely its locally obtained measurements. 
Following the rationale of the  majority of greedy algorithms (see, for example, \cite{Tr04,DTDS09,NeedTro09,DaMi09,NV09,HM11,PEE12})
  decisions concerning the support set should be based on a signal proxy of the form  $$\boldsymbol{A}_k^{T} \left(\boldsymbol{y}_k -\boldsymbol{A}_k \widehat{\boldsymbol{h}}_k\right)\approx \boldsymbol{A}_k^{T}\boldsymbol{A}_k \left(\wo_*-\widehat{\boldsymbol{h}}_k\right),$$ 
where $\widehat{\boldsymbol{h}}_k$ stands for the estimate of $\wo_*$ obtained at note $k$.
It is important to note that for such a proxy, the algorithm at every iteration, selects distinct indices. \textcolor{black}{Since the vast majority of greedy algorithm are iterative and hence it is important not to consider indices that have been previously picked. This is ensured by orthogonalizing the residual.} An enhanced proxy, which would allow multiple selection of indices, has the following form:
$$\supp_{s\text{ or }2s}\left(\boldsymbol{A}_k^{\textcolor{black}{T}}    \left(\boldsymbol{y}_k -\boldsymbol{A}_k \widehat{\boldsymbol{h}}_k\right)\right)\cup \supp_{s}\left(\widehat{\boldsymbol{h}}_k\right),$$
and in such cases, the estimation step is restricted to a set of indices larger than $s$. 
However, adopting such philosophies for the proxy signals in distributed learning leads to problems.  In distributed learning, the only quantity, which commonly affects all nodes, is 
the true vector $\wo_*$ ({T}his can be readily seen from \eqref{linsys}\textcolor{black}{).} Both   previously mentioned proxy
 signals  
 carry time--varying information from local estimates, which are different from node to node,
 and hence may pose problems to the  support set consensus, \textit{i.e.,} the nodes'
 ``agreement'' to the same support set.
 Thus, it is preferable to find a proxy which is close to $\wo_*$ instead of $\left(\wo_*-\widehat{\boldsymbol{h}}_k\right)$.  This is an important point for our analysis to follow, since such a mechanism will lead to the required consensus among the nodes.
 
To this direction, we adopt a similar methodology to the one   presented in \cite{Fouc11}, which selects the $s$ dominant coefficients 
of the following proxy:
\begin{equation}
 \hat{\boldsymbol{h}}_k+ \bm{A}_k^T(\bm{y}_k-\bm{A}_k\hat{\boldsymbol{h}}_k)
 \approx   \boldsymbol{h}_*.
\label{Eq:proposed_proxy}
\end{equation} 
The intuition of this proxy selection \textcolor{black}{is}  that the left hand side (lhs)
of \eqref{Eq:proposed_proxy} is a gradient descent iteration using the measurements $\bm{A}_k, \ \bm{y}_k$. It has been verified experimentally (see for example \cite{Fouc11} for the centralized case) that a few iterations are enough to recover the  support \textcolor{black}{under a certain condition on the restricted isometry constant}.

In decentralized/distributed learning, the subset selection and greedy update steps
can be further \textit{enriched} by exploiting the 
information that is available at the nodes of the network. Recall that,
 each node is able to 
 communicate and exchange data
  with the neighboring nodes, where a direct link is available.  
 Hence, the  signal proxies can be computed in a cooperative fashion
 and instead of exploiting only local quantities of interest, spatially received data
 will take part in the support set identification.
 As it will become clear in the simulations section, this turns out to be beneficial
 to the performance of the proposed algorithms.

In the sequel, we present a unified method for sparsity--aware distributed learning, adopting the greedy viewpoint.
 The  {core} steps, through which the algorithms for batch and adaptive learning ``spring'',  {are summarized below}:
\begin{itemize}
\item \textbf{Information Enhancement for   Subset Selection:} Each node exchanges information with its  neighbours, in order to construct the signal proxy, instead  of relying only on its local measurements. 
\item \textbf{Cooperative Estimation Restricted on the Support Set:} At each node,  a Least Squares is performed restricted on the support sets that were computed cooperatively in the previous step.
The neighbourhood exchanges their estimates and fuse them under  a certain protocol.
\item \textbf{Pruning Step:} Neighbouring nodes may have identified different support sets, 
due to the dissimilarity of their measurements. Hence, the result of the previously mentioned fusion may not give an  $s$--sparse estimate. To this end, pruning ensures that,
at  each iteration, an $s$--sparse vector  is  produced. 
\end{itemize}
 
Before we proceed any further, let us explain  the basic principles that underlie the cooperation among the nodes. Each node, at each iteration step, fuses under a certain rule the estimates and/or the measurements, which are  received from the neighbourhood.   
	 This fusion is dictated by the combination coefficients, which obey  the following rules:  $a_{r,k}  > 0$, if $r \in \N_k$, $a_{r,k} = 0$, 
	 if $r \notin \N_k$ and $\sum_{r\in\N_k}a_{r,k}=1, \forall k\in\N$.   Furthermore, it is assumed that each node
	 is a neighbor of itself, \textit{i.e.,} $a_{k,k}>0, \ \forall k\in\mathcal{N}$.
	 In words, every node assigns a weight to the estimates and/or measurements of the
neighborhood and a convex combination of them  is computed; the resulting aggregate is used properly in the learning scheme.
Two well known examples of combination coefficients are:  a) the Metropolis rule,  where 
\begin{equation*}
a_{r,k}=
\begin{cases}
\frac{1}{\mathrm{max}\left\lbrace\vert\N_k\vert,\vert\N_r\vert\right\rbrace}, & \text{if $r\in\N_k$ and $r\neq k$},\\
1 - \sum_{r\in\N_k \setminus k}a_{r,k}, & \text{if $r=k$},\\
0,& \mathrm{otherwise},
\end{cases}
\end{equation*}
and b) the uniform rule, in which:
\begin{equation*}
a_{r,k}=
\begin{cases}
\frac{1}{\vert\N_k\vert}, & \text{if $r\in\N_k$},\\
0,& \mathrm{otherwise}.
\end{cases}
\end{equation*}

\subsection{Batch  {Mode Learning} via the DiHaT algorithm}
In this section, a batch parsity-promoting algorithm, which we will call Distributed Hard Thresholding Pursuit Algorithm (DiHaT), is presented. The algorithmic steps are summarized in Table I.
\begin{table}[t]
    \renewcommand{\arraystretch}{2.1}
\caption{The Distributed Hard Thresholding  Algorithm (DiHaT)}   
   \resizebox{\columnwidth+0.21cm}{!}{
    \label{tab:spaMP}\vspace{-.2cm}
    \centering
    { \Huge
        \begin{tabular}[t]{clrc}
            \hline
            \multicolumn{3}{c}{Algorithm description} & \text{Complexity}\\
            \hline
            \multicolumn{2}{l}{$\boldsymbol{h}_{k,0}=\boldsymbol{0}_m,\quad \overline{\boldsymbol{y}}_{k,0} =\boldsymbol{y}_k, \quad \overline{\boldsymbol{A}}_{k,0} =\boldsymbol{A}_k, \quad\mathcal{S}_{k,0}=\emptyset, \quad \mathrm{sparsity \ level} \  \ \ s$} & \text{\{Initialization\}}\\
			\multicolumn{4}{l}{{\bf Loop}}\\             
            1: & \hspace{.3cm} $\overline{\boldsymbol{y}}_{k,n}=\sum\limits_{r\in\mathcal{N}_{k}}a_{r,k}\overline{\boldsymbol{y}}_{r,n-1}$ & \text{\{Combine local output measurements\}} & $\mathcal{O}(\vert\N_k\vert l)$ \\
            2: & \hspace{.3cm} $\overline{\boldsymbol{A}}_{k,n}=\sum\limits_{r\in\mathcal{N}_{k}}a_{r,k}\overline{\boldsymbol{A}}_{r,n-1}$ & \text{\{Combine local input measurements\}} & $\mathcal{O}(\vert\N_k\vert m l)$ \\
            3: & \hspace{.3cm} ${\mathcal{S}}_{k,n}=\mathrm{supp}_{s}\left(\boldsymbol{h}_{k,n-1}+\overline{\boldsymbol{A}}_{k,n}^T(\overline{\boldsymbol{y}}_{k,n}-\overline{\boldsymbol{A}}_{k,n}\boldsymbol{h}_{k,n-1})\right)$ & \text{\{Identify $s$ largest components\}} &  $\mathcal{O}(ml^2)$\\
            4: & \hspace{.3cm} $\hat{\boldsymbol{h}}_{k,n}=\arg\min_{\boldsymbol{h}}\left\lbrace \Vert \overline{\boldsymbol{y}}_{k,n}-\overline{\boldsymbol{A}}_{k,n} \boldsymbol{h}\Vert_{\ell_2}^2  \right\rbrace, \mathrm{supp}(\boldsymbol{h})\subset {\mathcal{S}}_{k,n}$ &\text{\{Signal Estimation\}} & $\mathcal{O}(m^2 l)$ \\
            5: & \hspace{.3cm} $\tilde{\boldsymbol{h}}_{k,n}=\sum\limits_{r\in\mathcal{N}_{k}}b_{r,k}\hat{\boldsymbol{h}}_{r,n}$ & \text{\{Combine local estimates\}} & $\mathcal{O}(\vert\N_k\vert m)$\\
            6: & \hspace{.3cm} $ {\boldsymbol{h}}_{k,n}=\mathrm{supp}_s(\tilde{\boldsymbol{h}}_{k,n})$ & \text{\{Identify $s$ largest components\}} & $\mathcal{O}(m)$\\
            
            \multicolumn{4}{l}{{\bf Until} halting condition is {\it true}}\\
            \hline
        \end{tabular}}
        }
\end{table}
In steps 1 and 2 {of Table I}, the nodes exchange their input--output measurements and
fuse them under a certain protocol. This information fusion is
dictated by the combination weights $a_{r,k}, \ \forall k\in\N, \ \forall l\in\N_k$.
It is shown later, that if these coefficients are chosen properly,
then $\overline{\bm{A}}_{k,n}$ and $\overline{\bm{y}}_{k,n}$  tend asymptotically to the average
values, \textit{i.e.,}  $\frac{1}{N}\sum_{r=1}^N\boldsymbol{A}_{r}$ and $\frac{1}{N}\sum_{r=1}^N\boldsymbol{y}_{r},$ respectively.
This   improves  significantly the performance of the algorithm, since
as the number of iteration increases, the information  related to the support set is accumulated, 
in the sense that   
 the support set identification and the parameter estimation procedure
will
contain information  which comes  from the entire network.
In step 3, the $s$ largest in amplitude coefficients of the signal
proxy are selected, and step 4 performs a Least--Squares 
operation restricted on   the support set, which is computed in
the previous step.
Next (step 5), the nodes exchange their estimates
and fuse them in a similar way as in steps 1, 2. Notice that, the combination coefficients
may be different from the weights used in input/output measurement combination (denoted as $b_{r,k}$). Finally, since the nodes have access to  different observations,
especially in the first iterations in which their input--output data
are not close to the average values, the estimated support
sets among the neighborhood may be different. This implies
that in step 5 there is no guarantee that the produced estimate
will be $s$--sparse. For this reason, in step 6, a thresholding operation
takes place and the final estimate at each node is $s$--sparse.
A proof concerning the convergence of the DiHaT algorithm can be found in Section V.

\begin{remrk}
\textcolor{black}{The DiHaT algorithm requires the transmission of $m\times l, l, m$ coefficients
from each node, coming from the input matrix, the output and the estimate, respectively.
There are two working scenarios to relax the required bandwidth for such an exchange. The exchange of the information takes place in blocks, in line with the available bandwidth constraints. Since this is a batch processing mode of operation, such a scenario will slow down the overall execution time, and has no effect on processing/memory resources.  The alternative scenario is to transmit only the obtained estimated, which amounts to $m$ coefficients per iteration step; as simulation show this amounts to a small performance degradation.
}
\end{remrk}
\section{THE ONLINE LEARNING SETUP}
 This section deals with an online learning version for the solution of the task.
 More specifically, our goal is to estimate a sparse unknown vector, $\wo_*\in \R^m$,
	exploiting an \textit{infinite} number of \textit{sequentially} arriving measurements collected at the $N$ nodes of an ad--hoc network \textit{e.g.}, \cite{LoSa08,CatSa10,Theo15}. 
	Each node $k$, at each  (discrete) time \textcolor{black}{instant},  
	has access to the   measurement  pair $(y_k(n),\uo_k(n))\in\R\times\R^m$,
	 which are related via the linear model: 
	\begin{equation}
	y_{k}(n)=\uo_{k}^T(n)\wo_*+v_k(n), \ \forall k\in \N, \ \forall n\in\Z,
	\end{equation}
	where the term $v_k(n)$ stands for the additive noise process at each node.
 
     In the literature, for  the adaptive distributed learning task, the following modes of cooperation have been proposed: 
	\begin{enumerate}
	\item \textbf{Adapt Then Combine (ATC) \cite{CatSa10}:} In this strategy, each node computes
	an intermediate estimate by exploiting locally sensed measurements. 
	After this step, each network agent receives these estimates from the neighbouring nodes
	and combines them, in order to produce the final estimate.  
	\item \textbf{Combine Then Adapt (CTA) \cite{LoSa08,ChoSlTh12}:} Here, the combination step precedes the adaptation one.
	\item \textbf{ADMM Based \cite{ScMaGi09}:} In this category, the computations are made in
	parallel and there is no clear distinction between the combine and
	the adapt steps.
	\end{enumerate}
    The ATC and the CTA cooperation strategies belong to the family of the diffusion--based algorithms. 
    In these schemes,  the cooperation among the nodes can be summarized as follows:
	each node receives estimates from the neighbourhood, 
	and then takes a convex combination of them.	
	In this paper, the ATC   strategy is followed, since it has been verified
	that it converges faster to a lower steady state error floor
	compared to the other methodologies \cite{TuSa12}.

  \subsection{Greedy--Based Adaptive Distributed Learning}
Our goal here is to 
reformulate properly the distributed greedy steps, which are 
described in Section \ref{sec:DiGreeLe}, so as to employ them in an adaptive scenario.
For simplicity, let us first  discuss the non--cooperative case.
In the batch learning setting, the proxy signal is constructed via the available measurements
at each node; \textit{i.e.,} the local measurements and the information received \textcolor{black}{from} the
neighbourhood.  As we have already described, in online learning the observations are
received sequentially, one per each time step and, consequently, a different route has
to be followed. 
Recall the definition of the signal proxy, given in \eqref{Eq:proposed_proxy}.
 A first approach is to make the following substitutions $\bm{A}_k^T\bm{y}_k$ with   $\bm{a}_k(n)y_k(n)$ and $\bm{A}_k^T\bm{A}_k$ with $\bm{a}_k(n)\bm{a}_k^T(n)$. A
 drawback of this choice is that the proxy is constructed exploiting just 
 a single pair of data, which in practice carries insufficient information. 
  Another viewpoint, which is followed here, is to rely on approximations 
  of the expected value of the previous quantities, \textit{i.e.,}  
  $\mathbb{E}[\bm{a}_k(n)y_k(n)]$, $\mathbb{E}[\bm{a}_k(n)\bm{a}_k^T(n)]$.
  Since the statistics are usually unknown {and might exhibit time--variations}, we rely on the following approximations,
   $\mathbb{E}[\bm{a}_k(n)y_k(n)]\approx \sum_{i=1}^n \zeta^{n-i} \bm{a}_k(i)y_k(i)$,
  $\mathbb{E}[\bm{a}_k(n)\bm{a}_k^T(n)]\approx\sum_{i=1}^n \zeta^{n-i}
   \bm{a}_k(i)\bm{a}_k^T(i)$,
  where $\zeta\in(0,1]$ is the so--called forgetting factor, incorporated in order to give 
  less weight to past values.
 
  The modified proxy, which is suitable for adaptive operation takes the following form:
\begin{align}\label{Eq:proposed_proxyAd}
	\bm{h}_k(n-1)+\tilde{\mu}_k(\overline{\boldsymbol{p}}_{k}(n)-\overline{\boldsymbol{R}}_{k}(n)\bm{h}_k(n-1))\approx \wo_*,
\end{align}
where $\boldsymbol{h}_k(n-1),  \  \overline{\boldsymbol{p}}_{k}(n)$ and $\overline{\boldsymbol{R}}_{k}(n)$
are defined in Table II and is $\tilde{\mu}_k$  the step size.
  The proposed proxy constitutes a distributed exponentially weighted extension of its centralized form (addressed in \cite{Fouc11}). Notice that   \eqref{Eq:proposed_proxyAd} defines a gradient descent iteration using the approximate statistics
$\overline{\boldsymbol{p}}_{k}(n)$ and $\overline{\boldsymbol{R}}_{k}(n)$. Assuming that we have
at our disposal the true statistical values, \textit{i.e.,}  $\overline{\boldsymbol{p}}_k:=\sum_{r\in\N_k}a_{r,k}{\boldsymbol{p}}_r$ and
$\overline{\boldsymbol{R}}_k:=\sum_{r\in\N_k}a_{r,k}{\boldsymbol{R}}_r$, then the recursion \eqref{Eq:proposed_proxyAd} 
converges to the true solution, \textit{e.g.,} \cite{sayedarxiv}. In practice, we use the 
previously mentioned approximate values, since we do not have access to the true statistics; as it will be become apparent in the simulations section, adopting such an approximation has little effect on the comparative performance of the algorithm.

\begin{table}[t]
 \renewcommand{\arraystretch}{2.3} 
 \caption{The Greedy Diffusion LMS Algorithm}
 \resizebox{\columnwidth+0.5cm}{!}{    
    \label{tab:spaMP2}\vspace{-.2cm}
    \centering
    {\Huge
        \begin{tabular}[t]{clrc}
			\hline
            \multicolumn{3}{c}{Algorithm description} & \text{Complexity}\\
            \hline
            \multicolumn{2}{l}{$\boldsymbol{h}_k(0)=0,\boldsymbol{p}_k(0)=0, \boldsymbol{R}_k(0)=0$} & \text{\{Initialization\}}\\
            \multicolumn{2}{l}{$0<\zeta\leq1$} & \text{\{Forgetting factor\}}\\
            \multicolumn{2}{l}{$0 < D$} & \text{\{Threshold for proxy selection\}}\\
            \multicolumn{3}{l}{{\bf For } $n:=1,2,\ldots $ {\bf do}}\\
            1: & \hspace{.3cm} $\boldsymbol{p}_{k}(n)=\frac{n}{n+1}\zeta\boldsymbol{p}_{k}(n-1)+ \frac{\boldsymbol{a}_{k}(n)y_{k}(n)}{n+1}$ & \hspace{-.3cm}\text{\{Form cross-correlation\}} & $O(m)$\\
            2: & \hspace{.3cm} $\overline{\boldsymbol{p}}_{k}(n)=\sum\limits_{l\in\mathcal{N}_{k}}b_{l,k}\boldsymbol{p}_{l}(n)$ & \text{\{Combine cross--correlation\}} & $O(\vert\mathcal{N}_k\vert m)$\\
 3: & \hspace{.3cm} $\boldsymbol{R}_{k}(n)=\frac{n}{n+1}\zeta\boldsymbol{R}_{k}(n-1)+ \frac{\boldsymbol{a}_{k}(n)\boldsymbol{a}_{k}^{T}(n)}{n+1} $ & \text{\{Form autocorrelation\}} & $O(m^2)$\\
            4: & \hspace{.3cm} $\overline{\boldsymbol{R}}_{k}(n)=\sum\limits_{l\in\mathcal{N}_{k}}b_{l,k}\boldsymbol{R}_{l}(n)$ &   		\text{\{Combine autocorrelation\}} & $O(\vert\mathcal{N}_k\vert m^2)$ \\
            \multicolumn{4}{l}{\quad\quad\quad{\textbf{If}} \ $\Vert \boldsymbol{h}_k(n-1)\Vert_{\ell_2}\leq D$}\\
            5: & \hspace{.3cm} $\begin{aligned}\widehat{\boldsymbol{S}}_{k,n}&=\supp_{s}\Big(\boldsymbol{h}_k(n-1)+\tilde{\mu}_k\Big(\overline{\boldsymbol{p}}_{k}(n)\\&-\overline{\boldsymbol{R}}_{k}(n)\boldsymbol{h}_k(n-1)\Big)\Big)\end{aligned}$ & \text{\{Identify large components\}}& $O(m)$ \\
             \multicolumn{4}{l}{\quad\quad\quad{\textbf{Else}}}\\
            6: & \hspace{.3cm} $\begin{aligned}\widehat{\boldsymbol{S}}_{k,n}&=\supp_{s}\Big(\frac{\boldsymbol{h}_k(n-1)}{\Vert \boldsymbol{h}_k(n-1)\Vert_{\ell_2}}+\tilde{\mu}_k(\overline{\boldsymbol{p}}_{k}(n)\\&-\overline{\boldsymbol{R}}_{k}(n)\frac{\boldsymbol{h}_k(n-1)}{\Vert \boldsymbol{h}_k(n-1)\Vert_{\ell_2}})\Big)\end{aligned}$   & \text{\{Identify large components\}}&  $O(m)$\\

			 7: & \hspace{.3cm} $\begin{aligned}\boldsymbol{\psi}_{k}(n)&=\boldsymbol{h}_{k|\widehat{\mathcal{S}}_{k,n}}(n-1)+\\ &\mu_k\boldsymbol{a}_{k|\widehat{\mathcal{S}}_{k,n}}(n)[y_{k}(n)-\boldsymbol{a}_{k|\widehat{\mathcal{S}}_{k,n}}^{T}(n)\boldsymbol{h}_{k|\widehat{\mathcal{S}}_{k,n}}(n-1)]
			 \end{aligned}$ &\text{\{LMS iteration\}} &  $O(s)$\\
            8: & \hspace{.3cm} $\widetilde{\boldsymbol{h}}_{k}(n)=\sum\limits_{r\in\mathcal{N}_{k}}c_{r,k}\boldsymbol{\psi}_{r|\widehat{\mathcal{S}}_{l,n}}(n)$ & \text{\{Combine local estimates\}} &  $O(\vert\mathcal{N}_k\vert s)$\\
            9: & \hspace{.3cm} $\widetilde{\mathcal{S}}_{k,n}=\supp_s\left(\widetilde{\boldsymbol{h}}_{k}(n)\right), \ \boldsymbol{h}_{k|\widetilde{\mathcal{S}}^{c}_{k,n}}(n)=\boldsymbol{0}$ & \text{\{Obtain the pruned support\}} & $O(m)$ \\
             
                    \hline
        \end{tabular}}
        }
\end{table}
Table \ref{tab:spaMP2} summarizes the core steps of the proposed Greedy Diffusion LMS (GreeDi--LMS). Steps 1--5 constitute the \textit{distributed greedy subset selection} process. It is worth pointing out that the Steps 1--4 create, in a 
cooperative manner, the basic elements needed for the proxy update and Steps 5--6 select the $s$--largest components.
Notice that  $D$ is a sufficiently large parameter, which is used to normalize the estimate
in the proxy computation step, if its norm is larger than this threshold, and it is employed
so as  to prevent the proxy from taking 
extensively large values.
 It should be pointed out that the \textit{distributed greedy update} is established via the Adapt--Combine LMS (Steps 7 \& 8) of \cite{CatSa10}. The diffusion Steps 2, 4 and 8 bring
the local estimates closer to the global estimates based on the entire network \cite{CatSa10}. Finally, because we are operating in a network of different nodes, some of them will achieve support set convergence faster than others. Therefore, at a node level, we pay more attention to the $s$--dominant positions via the introduction of a pruning step directly after the adaptation in the combined greedy update (Step  9).

\begin{remrk}
 The computational complexity of GreeDi--LMS is linear except from the update of $\boldsymbol{R}_k(n)$, which requires $\mathcal{O}(m^2)$  operations per iteration. However, exploitation of the shift structure in the regressor vector allows us to drop the computational cost to $\mathcal{O}(m)$, \textit{e.g.,} \cite{Say08}. Complexity as well as 
the number of transmitted coefficients, can be decreased if in steps 5 and 6
 each node exploits only local information.
\textcolor{black}{
Regarding the communication cost of the proposed algorithms, each node
transmits, overall $2*m+m^2$ values corresponding the estimation vector, the crosscorrelation vector and the autocorrelation matrix.
However, this cost can be  significantly decreased if one exploits the shift structure
of the regression vectors. More precisely, in that case, instead of the
$m^2$ entries of the autocorrelation matrix, only $2m$ coefficients need to be exchanged  
due to the Teoplitz structure of the autocorrelation matrix, 
\cite{Say08,Theo15}.
Finally, another way to bypass the need for transmission   of the autocorrelation matrices and the
crosscorrelation vectors is to compute them at node level. In particular, according 
to this scenario,
each node receives from the neighborhood the input/output data and computes
the respective quantities. Employing this methodology reduces the communication cost
to $2m+1$ coefficients, which correspond to the estimate and input vectors and the output.
This comes at the cost of an increased complexity.} 
 \end{remrk}

\begin{remrk}
 Another way to reduce complexity is by updating the gradient part of the proxy recursively{,} as it was done in \cite{MiBaKaTa10} for the centralized case. However, such an update, although it is computationally lighter, it results in slower support convergence, since the resulting exponentially weighted average contains all $\boldsymbol{h}_k(t), t:=1,\ldots, n-1$ and not only the current estimate. 
 \end{remrk}

 \section{Convergence Analysis}\label{Sec:performance_analysis}
 {In this section, the theoretical properties of the proposed schemes are studied.
 In a nutshell, we show that the DiHaT scheme converges, under certain assumptions
 regarding the noise vectors and a restricted isometry constant for the average
 value of the input matrices. Regarding the GreeDi--LMS we prove that the algorithm is
 {unbiased}, \textit{i.e.,} it converges in the mean to the true unknown vector. Finally,
 steady--state error bounds are provided.}
 
 \subsection{Convergence Analysis: The Batch Mode Learning Case}
\label{sec:ConAna}
The assumptions on the network and associated weights that are needed to establish convergence of the proposed algorithm are given next.
\begin{ass}
Let us define the $N\times N$ matrix $\boldsymbol{W}_1$, the coefficients formed by the combination weights $a_{r,k}$. This matrix satisfies the following assumptions:
\begin{enumerate}
\item $\boldsymbol{1}_N^T\boldsymbol{W}_1=\boldsymbol{1}_N^T$, where $\boldsymbol{1}_N\in \R^{N}$ is the vector of ones.
\item $\boldsymbol{W}_1\boldsymbol{1}_N=\boldsymbol{1}_N$.
\item $\lambda(\boldsymbol{W}_1-\boldsymbol{1}_N\boldsymbol{1}_N^T/N)<1$, where $\lambda(\cdot)$ stands for the maximum
eigenvalue of the respective matrix, \textit{ e.g.,} \cite{XiBo04}.
 \end{enumerate}
 \end{ass}
The same assumptions hold for   $\bm{W}_2$, with entries $b_{r,k}$.
Such assumptions are widely employed in distributed learning, \textit{e.g.,} \cite{sayedarxiv,XiBo04}.
Methods for constructing the combination matrix, so as to fulfil these assumptions in a decentralized fashion,
have been proposed, \textit{e.g.,} \cite{XiBo04}.\\
\textbf{Fact: \cite{XiBo04}} 
Under the above assumptions the following hold 
\begin{equation}
\lim_{n\rightarrow\infty} \overline{\boldsymbol{y}}_{k,n} = \overline{\boldsymbol{y}}:= \frac{1}{N}\sum_{r=1}^N\boldsymbol{y}_{r}, \ \forall k\in \N,
\end{equation}
as well as
\begin{equation}
\lim_{n\rightarrow\infty} \overline{\boldsymbol{A}}_{k,n} = \overline{\boldsymbol{A}} := \frac{1}{N}\sum_{r=1}^N\boldsymbol{A}_{r}, \ \forall k\in \N.
\end{equation}
The network nodes,  asymptotically, gain access to the mean value of the measurement matrix and measurement
vector by reaching consensus.\footnote{It should be pointed out that the theorem in \cite{XiBo04} considers the case
where each node has access and averages  a scalar. However, the results  can be readily generalized to the vector case (see for example \cite{CaYaMu09}). The
matrix case is treated in a similar way as in the vector one, if one vectorizes the matrix and combines the \textcolor{black}{resulting} vectors.} \\

\begin{ass}
\begin{enumerate}
\item \textcolor{black}{The noise terms $\bm{\eta}_k$ are assumed to be zero mean and of  bounded variance.
A typical example  is  the 
Gaussian distribution.}
 Moreover, they are spatially independent over the nodes.
\item For a large value of $n$, say $n_0$, we assume that \[\overline{\boldsymbol{y}}_{k,n} \approx\frac{1}{N}\sum_{r=1}^N{\boldsymbol{A}}_{r}\boldsymbol{h}_*  + \frac{1}{N}\sum_{r=1}^N \boldsymbol{\eta}_r \approx \overline{\boldsymbol{A}} \boldsymbol{h}_*, \ \forall n\geq n_0.\]
This is a direct consequence of the fact, that if the combination coefficients are chosen with respect to Assumptions 1.1-1.3, then 
$\lim_{n\rightarrow\infty} \overline{\boldsymbol{y}}_{k,n} = \frac{1}{N}\sum_{r=1}^N\boldsymbol{y}_{r}, \ \forall k\in \N$ and
$\lim_{n\rightarrow\infty} \overline{\boldsymbol{A}}_{k,n} = \frac{1}{N}\sum_{r=1}^N\boldsymbol{A}_{r}, \ \forall k\in \N$.

In other words, for   sufficiently large   $n_0$, in which the nodes have almost reached consensus, the noise contribution becomes equal  \textcolor{black}{to} 
  the average value of the individual noise terms.  Taking into consideration the Gaussianity of these  terms as well as the law of large numbers, \textit{e.g.,} \cite{PaPi02},
  for a large number of nodes, \textit{i.e.,} $N$, the average value of the individual noise terms vanishes out. 
\end{enumerate}
\end{ass}

\begin{theorem}
Assume that the matrix $\overline{\boldsymbol{A}}$ satisfies 
the following $3s$ Restricted Isometry Constant 
$\delta_{3s}<\frac{1}{3}$, $\forall k\in\N, \ \forall n\geq n_0$, \cite{Fouc11}.
Moreover, \textcolor{black}{suppose}  that Assumptions 1.1--1.3 and 2.1--2.2 hold true,
then:
\begin{align}
\Vert {\wot}_{n+1}-\wot_*  \Vert_{\ell_2} \leq \rho \Vert \underline{{\wo}}_{n}-\wot_*  \Vert_{\ell_2},
\end{align}
where $\rho=\max_{k\in\N}\left\lbrace\sqrt{\frac{8\delta_{3s}^{2}}{1-\delta_{2s}^{2}}}\right\rbrace < 1$,  
$\underline{\wo}_{n}=[\wo_{1,n}^T,\ldots,\wo_{N,n}^T]^T\in\R^{Nm}$ and  $\wot_*=[\wo_*^T,\ldots,\wo_*^T]^T\in\R^{Nm}$.\\
\end{theorem}

The proof is given in Appendix A.
 \subsection{Theoretical Analysis: The adaptive case}

As it is well known, \textcolor{black}{under some   assumptions} the LMS converges in the mean and
in the mean--squares sense to the true parameter vector.
This property is shared by the diffusion LMS schemes, \textit{e.g.,} \cite{CatSa10}.
Nevertheless, the diffusion--based sparsity--promoting LMS \cite{LoSa12} is not unbiased, due to the
convex regularization term  embedded in the cost function. 
In contrast GreeDi--LMS enjoys unbiasedness as demonstrated below.

\subsubsection{Convergence in the Mean}
\begin{ass}
\begin{enumerate}
\item The unknown vector $\wo_*$ is time-invariant, \textcolor{black}{the forgeting factor is
set to $1$}, and  
the regressors and the noise are ergodic processes, \textit{i.e.},
$\lim\limits_{n\rightarrow\infty} \frac{1}{n}\sum_{i=0}^n y_k(i)\bm{a}_k(i)=\bm{p}_k:=E[y_k(n)\bm{a}_k(n)]$ with probability \textcolor{black}{(w.p.) $1$} and
$\lim\limits_{n\rightarrow\infty}\frac{1}{n}\sum_{i=0}^n \bm{a}_k(i)\bm{a}^T_k(i)=\bm{R}_k:=E[\bm{a}_k(n)\bm{a}^T_k(n)]$ \textcolor{black}{w.p. $1$} , \textit{e.g.,} \cite{Say08,Hay96}.
\item The input is  a white noise sequence, \textit{i.e.,} $\bm{R}_k=\sigma^2_k\bm{I}_m, \ \forall k\in\N$.
\end{enumerate}
\end{ass}

\begin{lem}\label{lem:mean_covergence}
Assume that at every time \textcolor{black}{instant} we set $
\tilde{\mu}_k(n) = \frac{1}{\tilde{\nu}_k(n)}$, with ${\tilde{\nu}_k(n)}=\sum_{l\in\mathcal{N}_k}b_{r,k}\frac{1}{m}\sum_{i=1}^m [\bm{R}_l(n)]_{ii}$. Then, $\exists n_0\in \mathbb{Z}$ \textcolor{black}{(w.p.) $1$} such that $\widehat{\mathcal{S}}_{k,n}={\mathcal{S}}, \ \forall n\geq n_0, \ \forall k \in \mathcal{N}$, where
$\mathcal{S}$ is the support  set of the unknown vector.
\end{lem}
 The proof is given in Appendix B. 
 
%


The main conclusion drawn from Lemma  \ref{lem:mean_covergence} is that the true support can theoretically be obtained  in a finite number of steps. This is a direct consequence of the fact that the proxy converges asymptotically to a vector, which has the same support as the unknown one.  
 This has two direct implications, which can be explored further in order to reduce computational resources. First, once the signal proxy has converged then the algorithm
  of Table \ref{tab:spaMP2} becomes identical to the Adapt--Combine LMS presented in \cite{CatSa10}. Second, in stationary environments, if the signal proxy remains constant for a number of successive iterations we may   stop performing the proxy selection process of the algorithm so as to reduce complexity.
 
\begin{ass}
\begin{enumerate}
\item The input vectors are assumed to be independent of
the noise. 
\item (Independence): The  regressors are spatially and temporally independent.
This assumption allows us to consider the input vectors $\boldsymbol{x}_k(n)$
 independent of $\wo_k(j), \ \forall j\leq n-1$.
 Despite the fact that this 
 assumption is not true in
general, it is commonly adopted in 
adaptive filters, as it simplifies the analysis. 
\item The step--size at each node satisfies
\begin{equation*}
0<\mu_k<\frac{2}{\lambda_{\max}(\bm{R}_k)},
\end{equation*} 
\end{enumerate}
 
\end{ass}
\begin{theorem}\label{Thm:mean_covergence}
\textcolor{black}{Suppose} that Assumptions 3.1--3.2, 4.1--4.3 are true. 
Then, it holds that $\lim_{n\to\infty}\mathbb{E}[\wo_k(n)]=\wo_*$.
\end{theorem}
The proof is given in Appendix C.

It follows from Thm. \ref{Thm:mean_covergence} that the proposed algorithm avoids the major obstacle of (non--weighted) $\ell_{1}$--minimization methods, which cannot guarantee   recovery of the correct support and \textit{at the same time}  consistent estimation of nonzero  entries of $\wo_*$. 
In sparse $\ell_{1}$ regularized LMS--like algorithms, this is reflected by the introduction of a bias term in the mean converged vector.
 \subsection{Steady--State Error Bound}  
Our interest now turns into the $\ell_2$--norm of the error vector $\boldsymbol{h}_{k}(n)-\wo_*$ at each node in the  steady--state for the algorithm in Table II. 
  
 \begin{ass}
 
  For the proxy selection threshold,   $D$, we have that $D>2\Vert\wo_*\Vert_{\ell_2}+1$. Notice that knowledge of the exact value
of $\Vert\wo_*\Vert_{\ell_2}$ is not needed, since  $D$ can take any value larger than the previously mentioned threshold. This threshold is employed
  to avoid situations, in which extremely large values of the estimates occur, which are also carried in the 
proxy vector. 
 The next theorem is established in Appendix D.
 \end{ass}
\begin{theorem}\label{Thm:error_bound}  Each node $k\in\mathcal{N}$ produces an $s$--sparse approximation $\boldsymbol{h}_{k}(n)$ and 
the following asymptotic error bound for the whole network occurs:
\begin{align}
	\underline{\epsilon}(n)&\leq 2N\max_{k\in\mathcal{N}}\Big\{3(1-\mu_k\lambda_{min})+\sqrt{2}\delta_{3s,k}(\lambda,n\nonumber)\\&\left(1+\mu_k\lambda_{\max}\right)\Big\}\nonumber
	\sum_{k=1}^N \epsilon_{k}^{(2)}(n-1)\nonumber\\
	&\hspace{-1cm}+2N\max_{k\in\mathcal{N}}\left\{\sqrt{2}\sqrt{1+\delta_{2s,k}(\lambda,n)}\left(1+\mu_k\lambda_{\max}\right)\right\}\sum_{k=1}^{N}\Vert\overline{\boldsymbol{\eta}}_k(n)\Vert_{\ell_2}\nonumber\\
	&+2N\max_{k\in\mathcal{N}}\left\{\mu_k\|\boldsymbol{a}_{k|\Ss_{k,n}}^{T}(n)\|_{\ell_2}\right\}\sum_{k=1}^{N}|e_{o,k}(n)|\label{Eq:error_bound}
	\end{align}
where $\underline{\epsilon}(n):=\Vert {\underline{\wo}(n)}-\underline{\wo}_*  \Vert_{\ell_2}$, $\underline{\wo}(n)=[\wo_1^T(n),\ldots,\boldsymbol{h}_{N}^T(n)]^T\in\R^{Nm}$, $\underline{\wo}_*=[\wo_*^T,\ldots,\wo_*^T]^T\in\R^{Nm}$ 
and $\epsilon_{k}^{(2)}(n):=\|\boldsymbol{h}_{k}(n)-\wo_*\|_{\ell_2}$.
Furthermore, $e_{o,k}(n)$ is the estimation error of the optimum Wiener filter, $\delta_{3s,k}(\lambda,n)$ is the $3s$th order exponentially--weighted restricted isometry constant of the
measurement matrix \cite{MiBaKaTa10}, $\lambda_{k}$ is the maximum eigenvalue of the input covariance matrix $\boldsymbol{R}_k$. Finally, 
 the \textit{proxy noise disturbance} $\overline{\boldsymbol{\eta}}_k(n)$ results from the exponentially--weighted combination of nearby cross--correlation vectors with definition:
\begin{align*}
	\boldsymbol{p}_{k}(n)&=\sum_{r\in\mathcal{N}_{k}}a_{r,k}\left[\sum_{t=1}^{n}y_r(t)\boldsymbol{a}_r(t)\right]\\
 &=\sum_{r\in\mathcal{N}_{k}}a_{r,k}\left[\boldsymbol{R}_r(n)\wo_*+\frac{1}{n}\sum_{t=1}^{n}\boldsymbol{v}_{r}(t)\boldsymbol{a}_r(t)\right]\\
 &=\sum_{r\in\mathcal{N}_{k}}a_{r,k}\left[\boldsymbol{R}_r(n)\wo_*+\boldsymbol{\eta}_{r}(n)\right]=\overline{\boldsymbol{R}}_k(n)\wo_*+\overline{\boldsymbol{\eta}}_{k}(n).
\end{align*}
 \end{theorem}
The first and third term on the right hand side of Eq. \eqref{Eq:error_bound} reminds us the error bound of the Adapt--Combine LMS \cite{CatSa10}. The second term is analogous of the error bound of the Hard Thresholding Pursuit  algorithm, \cite{Fouc11}, restricted on the complement of the estimated support (corresponding to a batch of data of size $n$), mainly due to the exponentially weighted average proxy variant of \cite{Fouc11}. In a way, the second term remedies the pay--off of not having found completely the true support set after $n$ iterations.

\section{Computer Simulations}
\label{sec:simulations}
In this section, the performance of the DiHaT algorithm is studied in a batch scenario, where the nodes
of the network have access to a fixed number of measurements, whereas the performance
of the 
GreeDi--LMS  is tested in adaptive distributed examples.

\subsection{Performance Evaluation of the DiHaT}
\begin{figure*}[!t]
\graphicspath{{Figures/}}
\begin{minipage}[b]{.5\linewidth}
  \centering
\includegraphics[scale=0.6]{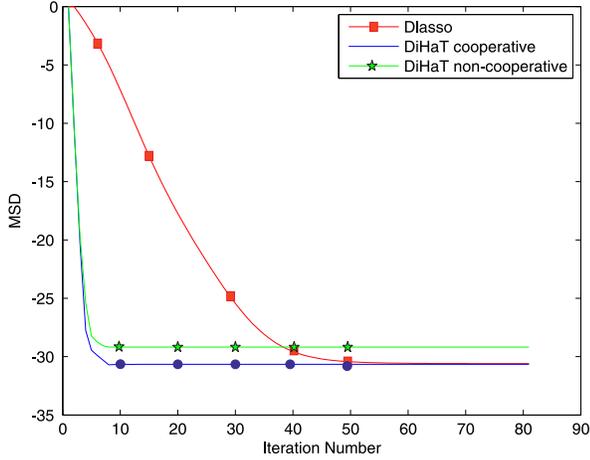}
  \centerline{(a)  MSD for the first experiment.\hspace{1.0 pt}}\medskip
\end{minipage}
\begin{minipage}[b]{.5\linewidth}
  \centering
\includegraphics[scale=0.6]{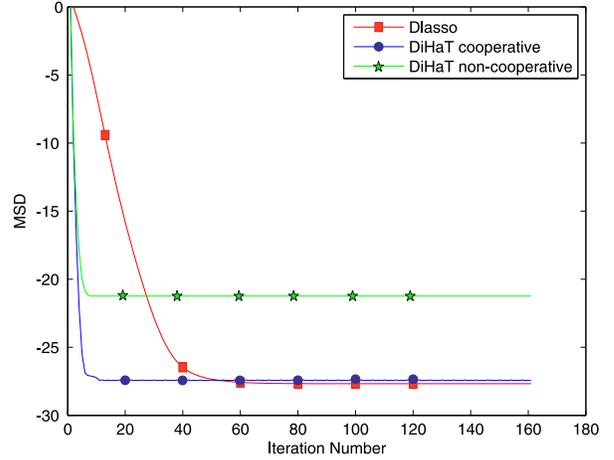}
  \centerline{ (b)  MSD for the second experiment.\hspace{1.0 pt}}\medskip
\end{minipage}
%
%
%
 \caption{Average MSD Performance of the DiHat Algorithm}
 \label{fig:res}
\end{figure*}

In the first experiment, the DiHaT is compared to the Dlasso proposed in   \cite{MaBaGi10}.
Moreover, the performance of the DiHaT is validated  in a scenario where the nodes
act autonomously with no cooperation; that is when they produce estimates relying
solely on their local input--output measurements.
A network with $N=20$ nodes is considered, the dimension of the unknown vector 
equals  $m=70$ and the number of measurements at each node is $l=55$. Furthermore,  $\Vert\bm{h}_*\Vert_{\ell_0}=10$. The coefficients of the 
input matrix $\bm{A}_{k}$ follow the Gaussian distribution
with zero mean and variance  $1$. Moreover, the noise is generated
with respect to the Gaussian distribution and a  Signal to Noise Ratio (SNR) at each node equal to 20 dB.
The combination coefficients $a_{r,k}, \ b_{r,k}$  are selected with respect to 
 the Metropolis rule \cite{XiBo04}. It is worth pointing out that, the combination 
 matrix   satisfies the properties 
 described in Section \ref{sec:ConAna}.
The performance metric  is the average normalized Mean--Square Deviation, which is given by
$\mathrm{MSD}(n)=\frac{1}{N}\sum_{k\in\N}\frac{\Vert\bm{h}_{k,n}-\bm{h}_*\Vert_{\ell_2}^2}{\Vert\bm{h}_*\Vert_{\ell_2}^2},$
and the curves result from  averaging of $100$ independent Monte Carlo (MC) runs. 
The first  computer experiment tests the proposed training based method versus Dlasso. The regularization parameter,  which is user defined in the Dlasso, is computed via a cross validation procedure, as proposed in   \cite{MaBaGi10}. Fig. \ref{fig:res}.a  illustrates  that the DiHaT outperforms the Dlasso, in the
sense that it converges faster to a similar error floor. Furthermore, the DiHaT in the non--cooperative
scenario converges to a higher error floor, which indicates that the cooperation among the 
nodes enhances the results.  It should be noticed  that, 
the Least Squares operation of the DiHaT takes place in the identified support set, which reduces significantly
the dimensionality, in contrast to the Dlasso, where all   operations take place in the original space
of dimension $m$. 

In the second experiment, the parameters remain the same as in the first one, but  now  
a lower sparsity level,  $\Vert\bm{h}_*\Vert_{\ell_0}=20$, is considered.
As it can be seen from Fig. \ref{fig:res}.b, the Dlasso outperforms significantly
the non--cooperative counterpart of the DiHat. Nevertheless, the enhanced  performance
of the cooperative DiHaT, compared to the Dlasso, is retained. Intuitively, the previously
mentioned behaviour  is a consequence of the fact that a larger support set is
more difficult to be identified, which can be seen from the performance degradation of the 
non--cooperative DiHaT. However, this problem can be overcome by identifying the support--set cooperatively, \textcolor{black}{which enhances the performance of the DiHaT algorithm as
  illustrated in  Fig. \ref{fig:res}.b.}

\begin{figure}
 \centering
  \includegraphics[scale=0.6]{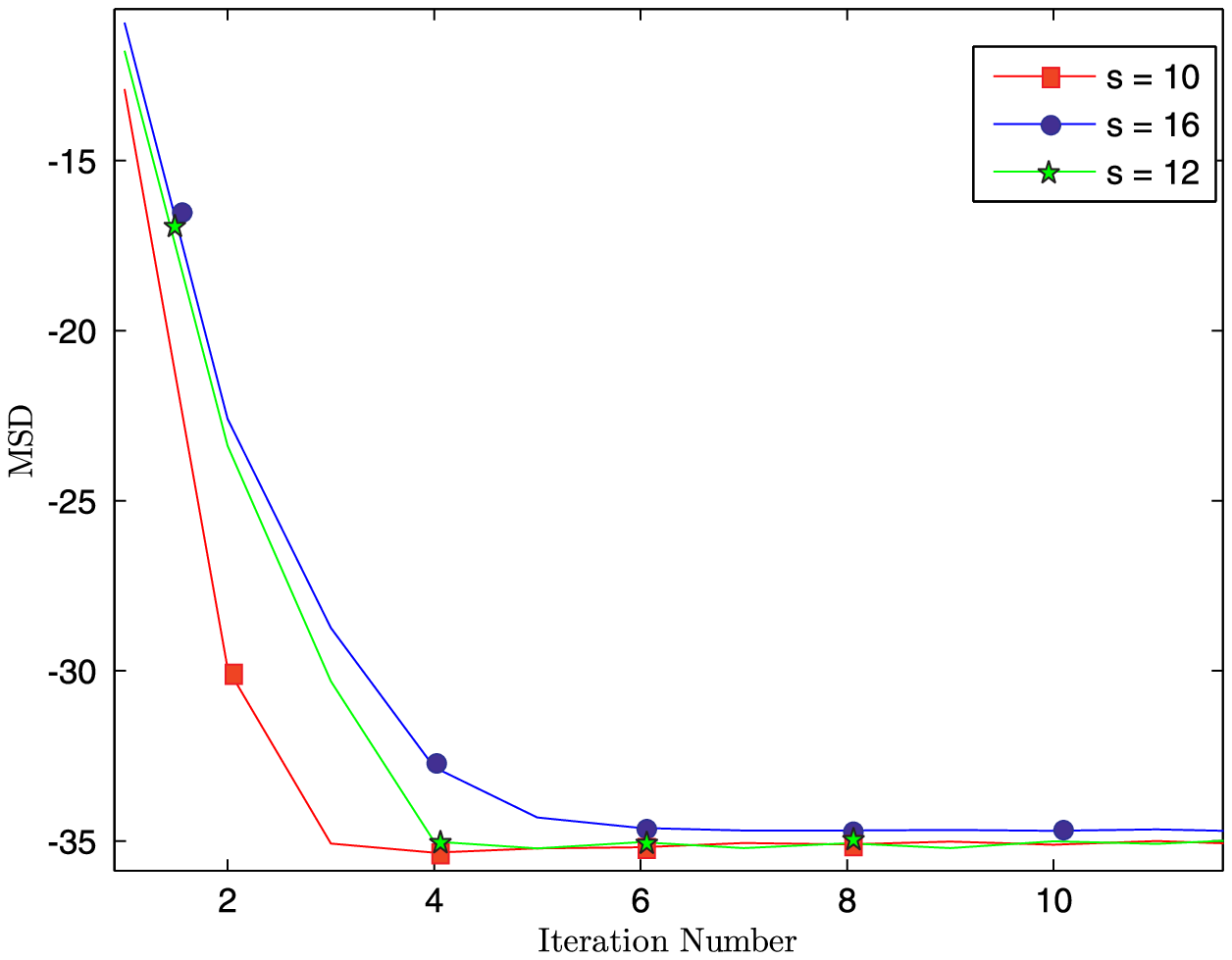}
 \label{exp1f}
  \caption{Sensitivity of the DiHaT to the $s$ parameter.}
 \label{exp1ff}
 \end{figure}

In the following experiments, we study the robustness of the DiHaT and the Dlasso algorithms, 
 by their sensitivity to non ``optimized'' configurations. First of all, it is worth pointing out
 that a single user--defined parameter is employed in our proposed scheme (the sparsity level  
 $s$), whereas in the Dlasso two user--defined parameters have to be tuned.
 The sensitivity of  DiHaT on $s$ is examined. Moreover, we study how different choices of a 
parameter, which will be denoted by $c$ and is 
 associated to the ADMM, affect the performance of the Dlasso.
From Fig. \ref{exp1ff}, it can be readily seen that the DiHaT is rather insensitive if one sets
the parameter $s$ to values larger than $\Vert\bm{h}_*\Vert_{\ell_0}$, which here equals to $10$. 
To be more specific, if $s=12$ 
then the algorithm converges slower, compared to the curve occurring by the optimized scenario, where $s=10$.
Furthermore, if $s=16$ then we observe a slower convergence speed and  a slightly higher
error floor. Nevertheless, in both cases, the performance degradation and, consequently,
 the sensitiveness
of the algorithm to the parameter $s$  is rather small.
Fig. \ref{exp2ff} illustrates the MSD curves of the Dlasso for the optimum choice of 
$c$, which equals to $0.3$, for $c=0.4$ and $c=0.5$. As it can be seen, 
the choice  of this parameter affects significantly the convergence speed.

 \begin{figure}
 \centering
  \includegraphics[scale=0.6]{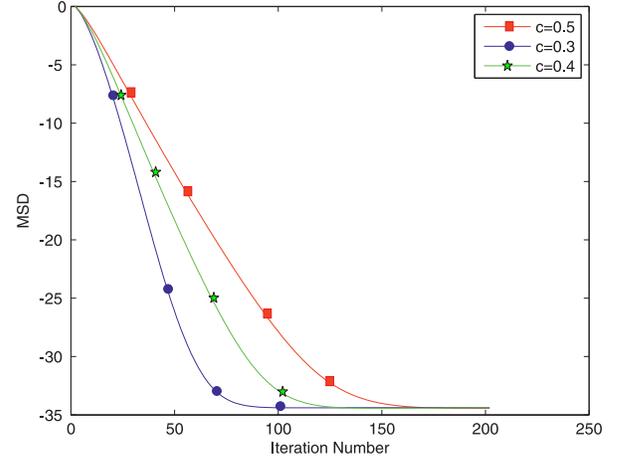}
 \label{exp2f}
  \caption{Sensitivity of the Dlasso to the $c$ parameter.}
 \label{exp2ff}
 \end{figure}
 
 \textcolor{black}{A drawback of the DiHaT compared to Dlasso  
 is that the former requires a sufficient number of observations at each node,
 whereas the latter   requires a certain  number of data to be spread
 throughout the network.   In particular, the DiHaT requires a RIP condition at node level,
 whereas the Dlasso does not make any assumptions regarding the number of measurements
 at each node.
 The goal of this experiment is twofold. The first goal   is to shed light on  the previously 
 mentioned issue, by evaluating the performance of the algorithms in a scenario
 where each node has access to a small number of measurements.
  The second one is to validate the performance of a variation of the DiHaT algorithm,
  in which the nodes exchange and fuse estimate vectors solely.
  This is of significant interest in applications where exchanging measurement data is not feasible,
  due to energy/privacy constraints and in scenarios where the number of measurements
  varies from node to node and data fusion is not possible.
 }
 
  \textcolor{black}{We consider the experimental set up of the first experiment where the
  number of measurements at each node equals   $15$. As it can be seen from
  Fig 5,   the non--cooperative greedy based algorithms fails to 
  estimate the unknown vector accurately, since the error floor is high.
  On the contrary, the fully cooperative DiHaT as well as the DiHaT with estimate exchange
  exhibit an enhanced performance compared to that of the non--cooperative algorithm. 
  Indeed, the full cooperative DiHaT converges relatively fast to a low error floor,
  whereas the DiHaT with estimate exchange converges to a significantly lower 
  error floor compared to the non--cooperative algorithm, but larger 
  compared to the DiHaT with data exchange. 
   Finally, the Dlasso outperforms the greedy--based algorithms as it converges
   to the lowest error.
    }

 \begin{figure}
 \centering
  \includegraphics[scale=0.6]{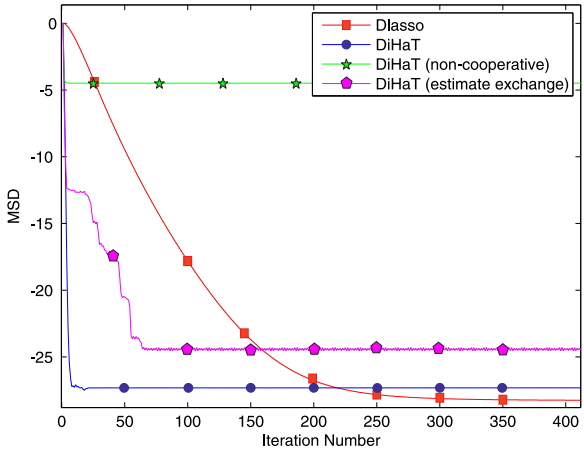}
 \label{exp2f}
  \caption{Performance evaluation in a scenario where the nodes
  have access to a small number of measurements.}
 \label{exp2ff}
 \end{figure}
\subsection{Performance Evaluation of the GreeDi LMS}
In this {sub}section, we present computer simulations {of the adaptive variant} of the proposed algorithm, {and compare it against} the sparsity promoting
diffusion LMS--based scheme   of \cite{LoSa12} (SpaDLMS).
The performance metric is the average Mean Square Deviation, which equals to 
 $\mathrm{MSD}(n)=1/N \sum_{k=1}^N\Vert \wo_{k}(n)-\wo_* \Vert^2$ and the 
 curves result from $100$   MC  runs.
  At each MC run, a new sparsity pattern is generated and the non--zero elements of the parameter $\wo_*$ for that run  are draws from a multivariate Gaussian distribution $\wo_{*|\mathcal{S}} \sim \mathcal{N} (0, \boldsymbol{I})$.

In the sixth experiment, we consider an ad--hoc   network \textcolor{black}{consisting} of $N=10$ nodes. The unknown vector has dimension equal to $m=100$, with $10$ non--zero coefficients ($10\%$ sparsity ratio ($s/m$)).
The input is drawn from a Gaussian distribution, with mean value equal to zero
and variance equal to $1$, whereas the    variance of the noise equals to $\sigma_k^2=0.01\eta_k \ \forall k\in\N$, 
where $\eta_k\in[0.5,1]$ is uniformly distributed.
The combination coefficients are chosen with respect to the Metropolis rule. 
In this experiment, it is assumed that both algorithms are optimized in the sense that the
regularization parameter used in \cite{LoSa12} is chosen according to the optimum rule presented in this study,  whereas in the proposed algorithm
we assume that we know the number of non--zero coefficients. 
\textcolor{black}{The   regularization function, which enforces sparsity,
 is chosen similarly to the one
proposed in \cite{LoBaSa13}.}
Finally, the step--sizes are chosen so that the algorithms exhibit a similar convergence speed,
and the forgetting factor $\zeta$ equals to $1$.
From Fig. 6, it can be seen that the proposed algorithm outperforms SpaDLMS significantly,
since it converges to a lower steady state error floor, at a similar  convergence speed.

 \begin{figure}
 \centering
  \includegraphics[scale=0.6]{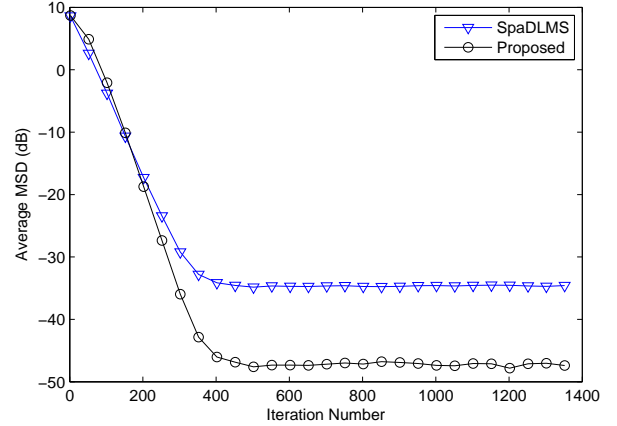}
 \label{projfig}
  \caption{Average MSD for the sixth experiment. \textcolor{black}{The stationary  scenario.}}
 \label{exp1}
 \end{figure}

\begin{figure}
\centering
 \includegraphics[scale=0.6]{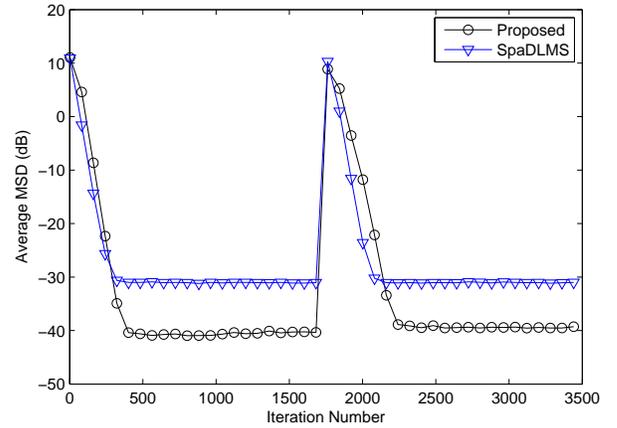}
\label{exp_ad3}
\caption{Average MSD for the seventh experiment. \textcolor{black}{The time varying scenario.}}
\label{exp3}
\end{figure}

In the seventh experiment, we examine the tracking behavior of the proposed scheme,
\textit{i.e.}, the performance of the proposed algorithm  in time--varying environments  
and the sensitivity in the case where our parameters are not optimized.
In order to achieve these two goals, the following scenario is considered. 
We assume that at the first 1450 iterations the parameters are the same as in the first
experiment, with the exception of the forgetting factor which now equals to
$\zeta=0.99$. At the next time instant   the channel undergoes a sudden change. Specifically, the
number of non--zero coefficients equals to $15$. Notice that, the sparsity level for the GreeDi LMS
is chosen equal to $15$, hence the parameter setting
is no more ``optimal''.
From Fig. \ref{exp3}, it can be readily seen that the proposed algorithm
enjoys a good tracking speed, since after the sudden change, it reaches at steady state, faster than the SpaDLMS.
 
\begin{figure}
 \centering
  \includegraphics[scale=0.6]{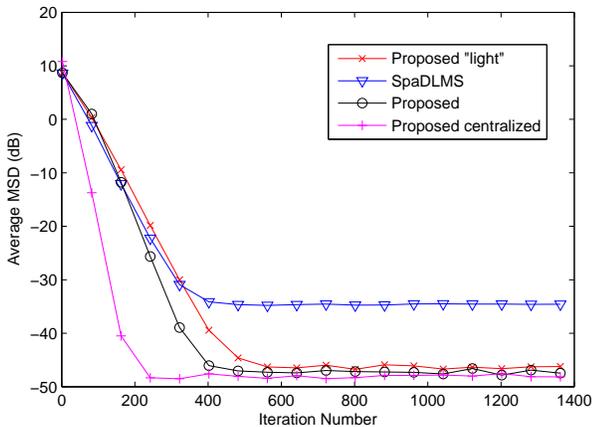}
 \label{projfig}
  \caption{Average MSD for the eighth experiment. \textcolor{black}{Performance evaluation of the ``light'' GreeDI LMS.}}
 \label{expadf}
 \end{figure}

Finally, in the eighth experiment, we study the performance of a modified version of the GreeDI LMS.
To be more specific, we assume that the proxy is computed in a similar way as in \cite{MiBaKaTa10} (see also Remark 3).
In that case, the proxy computation involves operations between vectors and the complexity drops to $O(m)$.
The parameters are the same as in the  fifth experiment, and this modified algorithm is compared to the
GreeDI LMS and the SpaDLMS. \textcolor{black}{Furthermore, we include the GreeDI LMS, operating in a centralized
mode, the performance of which serves as a benchmark.} The step--sizes where chosen so that the
decentralized algorithms exhibit a similar convergence speed.
Finally, the step--size for the centralized algorithm equals to that of the decentralized
one. 
Fig. \ref{expadf} illustrates, that the GreeDI LMS  outperforms
both the ``light'' version of the GreeDI LMS and the SpaDLMS. 
Furthremore, the ``light'' GreeDI LMS converges to a lower
error floor compared to the SpaDLMS. Finally, as it is expected,
the centralized algorithm outperforms all the decentralized ones,
since it exhibits a  faster convergence speed. 

\begin{remrk}
\textcolor{black}{It should be mentioned that the enhanced performance of the GreeDi LMS, compared to the
SpaDLMS, comes at the expense of a higher complexity, since the complexity
of the latter algorithm is of order $\mathcal{O}(m)$.  
Finally, the complexity of the 
``light'' version of the GreeDI LMS is $\mathcal{O}(m)$ and each node
transmits, at each time step, $2m$ coefficients, whereas the SpaDLMS requires
the transmission of $m$ coefficients.
}
\end{remrk}
 \section{Conclusions}
This work presented  greedy methods for  linear parameter estimation. Two operational modes were studied. Under the batch mode, where  each node has access to a finite number of measurements,
as well as on adaptive learning, where measurements arrive 
 sequentially and the estimates are updated
dynamically.     
 The behavior of both schemes was analyzed theoretically and via simulations. The conditions, under which the 
batch algorithm converges are given whereas the adaptive algorithm is shown to
converge, in the mean sense, and error bounds are derived.
  Future
research will be focused on sparsity promoting   algorithms for nonlinear
distributed systems.
\appendices
\section{Proof of Theorem 1}
Let us define the networkwise vectors 
$\underline{\hat{\wo}}_{n}=[{\hat{\wo}}_{1,n}^T,\ldots,{\hat{\wo}}_{N,n}^T]^T \in\R^{Nm}$ and 
$\underline{\tilde{\wo}}_{n}=[{\tilde{\wo}}_{1,n}^T,\ldots,{\tilde{\wo}}_{N,n}^T]^T \in\R^{Nm}$.
 Using lemma 4.5 of \cite{NeedTro09}, \textcolor{black}{which provides a bound for the pruned
 estimate,} 
we have that:
\begin{align}
	\Vert {\wot}_{n}-\wot_*  \Vert_{\ell_2} \leq 2 \Vert \underline{\tilde{\wo}}_{n}-\wot_*  \Vert_{\ell_2}.\label{Eq:pruned_w}
\end{align}

Next, let us define the $Nm\times Nm$ consensus matrix, $\bm{P}$, 
 which is defined as  $\bm{P}=\bm{W}_2\otimes \bm{I}_m$, where  the coefficients
of the matrix $\bm{W}_2$ constitute the weights $b_{r,k}$. Recall that $\bm{P}\widetilde{\wot}=\widetilde{\wot},\ \forall \widetilde{\wot}\in \mathcal{O}$, where 
$\mathcal{O}\subset\R^{Nm}$ is the consensus subspace with definition $\mathcal{O}:=\lbrace\widetilde{\wot}\in \R^{Nm}:\widetilde{\wot}=[\widetilde{\wo}^T,\ldots,\widetilde{\wo}^T]^T, \ \widetilde{\wo}\in\R^m \rbrace$. Moreover, recall that $\Vert\bm{P}\Vert = 1$ (see \cite{ChoSlaKoT12}) and hence
\begin{align}
\Vert \widetilde{\wot}_n-\wot_*  \Vert_{\ell_2} &= \Vert \bm{P} \underline{\hat{\wo}}_n - \wot_*  \Vert_{\ell_2} \nonumber\\&=
\Vert \bm{P} \underline{\hat{\wo}}_n - \bm{P}  \wot_*  \Vert_{\ell_2} \leq \Vert   \underline{\hat{\wo}}_n -   \wot_*  \Vert_{\ell_2}.
\label{eq1}
\end{align}
Fix a node, say $k$.
It holds that:
\begin{equation}
\Vert {\hat{\wo}}_{k,n}-\wo_*\Vert_{\ell_2}^2=\Vert ({\hat{\wo}}_{k,n}-\wo_*)_{\mathcal{S}_{k,n}}\Vert_{\ell_2}^2 +\Vert ({\hat{\wo}}_{k,n}-\wo_*)_{{\mathcal{S}^C}_{k,n}}\Vert_{\ell_2}^2 
\label{fouc:ineq1}
\end{equation}
\textcolor{black}{Next, we will exploit results from
 \cite[Theorem 3.5]{Fouc11}, where 
the distance between the estimate and the unknown vector  
restricted on the estimated support set or the complementary of this set,
is given.}
So, following similar steps as in \cite[Theorem 3.5]{Fouc11}, and taking into consideration Assumption 2.1, which states
that for $n\geq n_0$ the noise term is negligible, the error term  supported on the estimated support set satisfies
\begin{equation}
\Vert ({\hat{\wo}}_{k,n}-\wo_*)_{\mathcal{S}_{k,n}}\Vert_{\ell_2} \leq \delta_{2s}\Vert {\hat{\wo}}_{k,n}-\wo_*\Vert_{\ell_2}.
\label{fouc:ineq2}
\end{equation}
Moreover, according to \cite[Theorem 3.5]{Fouc11},   the second term of the right hand side in (\ref{fouc:ineq1})  satisfies
\begin{equation}
\Vert ({\hat{\wo}}_{k,n}-\wo_*)_{{\mathcal{S}^C}_{k,n}}\Vert_{\ell_2} \leq \sqrt{2} \delta_{3s} \Vert {{\wo}}_{k,n-1}-\wo_*\Vert_{\ell_2}. 
\label{fouc:ineq3}
\end{equation}
Combining (\ref{fouc:ineq1})-(\ref{fouc:ineq3}) we arrive at the following inequality
\begin{equation}
\Vert {\hat{\wo}}_{k,n}-\wo_* \Vert_{\ell_2} \leq \sqrt{\frac{2\delta_{3s}^{2}}{1-\delta_{2s}^{2}}}\Vert {{\wo}}_{k,n-1}-\wo_*\Vert_{\ell_2}. 
\label{fouc:ineq4}
\end{equation}
Taking into consideration (\ref{fouc:ineq4}), for the whole network we have
\begin{equation}
\Vert \underline{\hat{\wo}}_{n}-\underline{\wo}_* \Vert_{\ell_2} \leq \max_{k\in\N}\left\lbrace\sqrt{\frac{2\delta_{3s}^{2}}{1-\delta_{2s}^{2}}}\right\rbrace \Vert {\underline{\wo}}_{n-1}-\underline{\wo}_*\Vert_{\ell_2}. 
\label{fouc:ineq5}
\end{equation}

Now, combining inequalities \eqref{Eq:pruned_w}, \eqref{eq1}, (\ref{fouc:ineq5}) it follows
\begin{equation}
\Vert \underline{{\wo}}_{n}-\underline{\wo}_* \Vert_{\ell_2} \leq \max_{k\in\N}\left\lbrace\sqrt{\frac{8\delta_{3s}^{2(k,n)}}{1-\delta_{2s}^{2 (k,n)}}}\right\rbrace \Vert {\underline{\wo}}_{n-1}-\underline{\wo}_*\Vert_{\ell_2}. 
\label{fouc:ineq6}
\end{equation}
Since by definition $\delta_{2s}^{(k,n)}\leq \delta_{3s}^{(k,n)}$, \textit{e.g.,} (see for instance \cite{Fouc11}), and by assumption,  $\delta_{3s}^{(k,n)} < \frac{1}{3},$ $\forall k\in \N$ it follows  that $\max_{k\in\N}\left\lbrace\sqrt{\frac{8\delta_{3s}^{2(k,n)}}{1-\delta_{2s}^{2 (k,n)}}}\right\rbrace < 1$,
and the proof is completed.
 \section{Proof of Lemma 1}
 
Under Assumption 3, we have that $\lim\limits_{n\rightarrow\infty}[\bm{R}_k(n)]_{ii} = \sigma_k^2$ \textcolor{black}{(w.p.) $1$}\footnote{\textcolor{black}{All the results from now on, hold (w.p.) $1$. To avoid repetition   this statement is omitted.}}. So, recalling the definition of $\tilde{\nu}_k(n)$ we observe
\begin{align}
\lim\limits_{n\rightarrow\infty}\tilde{\nu}_k(n) &=  \lim\limits_{n\rightarrow\infty}\sum_{r\in\mathcal{N}_k}b_{r,k}\frac{1}{m}\sum_{i=1}^m [\bm{R}_r(n)]_{ii} \nonumber \\ &=  \sum_{r\in\mathcal{N}_k}b_{r,k}\frac{1}{m} \sum_{i=1}^m \lim\limits_{n\rightarrow\infty}  [\bm{R}_r(n)]_{ii} = \sum_{r\in\mathcal{N}_k}b_{r,k} \sigma_r^2.
\label{eqnu}
\end{align}
Now, since $\lim\limits_{n\rightarrow\infty} \overline{\bm{R}}_k(n) = \lim\limits_{n\rightarrow\infty}\sum_{r\in\mathcal{N}_k}b_{r,k}\bm{R}_r(n) = \sum_{r\in\mathcal{N}_k}b_{r,k} \sigma_r^2\bm{I}_m$ and from (\ref{eqnu}) it can be readily observed that 
\begin{equation}
\lim\limits_{n\rightarrow\infty} \tilde{\mu}_k(n) \overline{\bm{R}}_k(n) = \bm{I}_m.
\label{eqR}
\end{equation}
The  vector used for the proxy computation, if $\Vert \wo_k(n-1) \Vert_{\ell_2} \leq D$, can be rewritten as 
\begin{align}
&\wo_k(n-1)+\tilde{\mu}_k(n)(\overline{\bm{p}}_k(n)-\overline{\bm{R}}_k(n)\wo_k(n-1))\nonumber\\ &=
 \left(\bm{I}_m-\tilde{\mu}_k(n)\overline{\bm{R}}_k(n)\right) \wo_k(n-1)+\tilde{\mu}_k(n)\overline{\bm{p}}_k(n)
\label{eqli}
\end{align}
and if $\Vert \wo_k(n-1) \Vert_{\ell_2} > D$:
\begin{align}
&\frac{\wo_k(n-1)}{\Vert \wo_k(n-1) \Vert_{\ell_2}}+\tilde{\mu}_k(n)\left(\overline{\bm{p}}_k(n)-\overline{\bm{R}}_k(n)\frac{\wo_k(n-1)}{\Vert \wo_k(n-1) \Vert_{\ell_2}}\right)\nonumber \\
 &=
\left(\bm{I}_m-\tilde{\mu}_k(n)\overline{\bm{R}}_k(n)\right) \frac{\wo_k(n-1)}{\Vert \wo_k(n-1) \Vert_{\ell_2}}+\tilde{\mu}_k(n)\overline{\bm{p}}_k(n)
\label{eqli3}
\end{align}

Let us first study the case where $\Vert \wo_k(n-1) \Vert_{\ell_2} < D$. 
Taking into consideration  \eqref{eqR} and (\ref{eqli}) and the fact that $\Vert \wo_k(n-1) \Vert_{\ell_2} < D$, 
we have that 
\begin{align}
&\lim\limits_{n\rightarrow\infty} \left(\wo_k(n-1)+\tilde{\mu}_k(n)(\overline{\bm{p}}_k(n)-\overline{\bm{R}}_k(n)\wo_k(n-1))  \right)\\ &=\lim\limits_{n\rightarrow\infty}\tilde{\mu}_k(n)\overline{\bm{p}}_k(n)=\frac{1}{\overline{\sigma}_k^2}\overline{\boldsymbol{p}}_k,
\label{eqli2}
\end{align}
where $\overline{\sigma}_k^2:=\sum_{r\in\N_k}b_{r,k}{\sigma}_l^2$ and $\overline{\bm{p}}_k:=\sum_{r\in\N_k}b_{r,k}{\bm{p}}_l$.
Obviously, if $\Vert \wo_k(n-1) \Vert_{\ell_2} > D$ and if we  combine (\ref{eqR}) with (\ref{eqli3}), the same limit occurs.
From the Wiener--Hopf equation, e.g., \cite{Hay96}, it holds that
\begin{equation}
\frac{1}{\overline{\sigma}_k^2}\overline{\boldsymbol{p}}_k =\frac{1}{\overline{\sigma}_k^2}\sum_{r\in \mathcal{N}_k}{b_{r,k}}\boldsymbol{R}_{r} \boldsymbol{h}_*.
\label{wien}
\end{equation}
Nevertheless, from the diagonality of the matrices $\boldsymbol{R}_{k}, \ \forall k\in\N$ (Assumption 3.2), it  can be readily 
obtained from (\ref{wien}) that 
\begin{equation}
\mathrm{supp}\left(\frac{1}{\overline{\sigma}_k^2}\overline{\boldsymbol{p}}_k\right)=\mathcal{S}
\label{supps}
\end{equation}
{Equations \eqref{eqli2} and \eqref{supps} complete our proof, since
 the proxy converges to a vector, which has the same support set as $\wo_*$ and obviously,
 after a finite number of iterations the $s$--largest in magnitude coefficients of these two 
 vectors will coincide.}
 \textcolor{black}{Intuitively, the fact that the proxy converges to a vector, with the same
 support set as $\wo_*$ implies that these two vectors will be arbitrarily close $n\geq n_0$,
  for a sufficiently 
 large $n_0$. If the $s$ largest in amplitude coefficients of the proxy, do not coincide
 with the positions of the true support set, then a coefficient of the proxy, say in the
 position $i_0\notin\mathcal{S}$ will be larger in amplitude than a coefficient
 $i_0'\in\mathcal{S}$. This contradicts the fact that the previously mentioned vectors 
 will be arbitrarily close, since their distance will be \textit{at least larger} than
 the amplitude of the coefficient $i_0$ of the proxy. 
 }
 
 \section{Proof of Theorem 3}
 Notice that from the previous Lemma,  $\forall n\geq n_0, \ \forall k\in \N$ the algorithmic scheme drops to the
Adapt--Combine LMS presented in \cite{CatSa10}. 
Now, let us define the networkwise error vector $\tilde{\underline{\wo}}(n)=[\wo_1^T(n)-\wo_*^T,\ldots,\boldsymbol{h}_{N}^T(n)-\wo_*^T]^T\in\R^{Nm}$.
Using similar arguments as in \cite{CatSa10} we have that $\forall n \geq n_0$ 
the error vector for the whole network can be written as follows
\begin{align}
\tilde{\underline{\bm{h}}}(n)=\bm{P}\left(\tilde{\underline{\wo}}(n-1)-\mathcal{M}[\bm{D}(n)\underline{\wo}(n-1)+\bm{G}(n)]\right),
\label{eqC}
\end{align}
where $\mathcal{M}=\mathrm{diag}\lbrace\mu_1\bm{I}_m,\ldots,\mu_N\bm{I}_m\rbrace$,  $\bm{D}(n)=\mathrm{diag}\lbrace \boldsymbol{a}_1(n)\boldsymbol{a}_1^T(n),\ldots\,\boldsymbol{a}_N(n)\boldsymbol{a}_N^T(n)\rbrace$,\linebreak $\bm{G}(n)=[\boldsymbol{a}_1^T(n){v}_1(n),\ldots\,\boldsymbol{a}^T_N(n){v}_N(n)]^T$  and $\bm{P}$ is the $Nm\times Nm$ consensus matrix, which contains
the combination coefficients $c_{r,k}, \ \forall k\in \N, \forall r\in \N_k$, e.g., \cite{ChoSlaKoT12}.

Taking expectation in (\ref{eqC}) and taking into consideration the
assumptions we have that
\begin{align}
E[\tilde{\underline{\bm{h}}}(n)]=\bm{P}\left(\bm{I}_{Nm}-\mathcal{M}\mathcal{D}\right)E[\tilde{\underline{\bm{h}}}(n-1)], \ \forall n\geq n_0
\label{eqcit}
\end{align}
where $\bm{I}_{Nm}$ is the identity matrix of dimension $Nm\times Nm$ and $\mathcal{D}:=\mathrm{diag}\lbrace\bm{R}_1,\ldots,\bm{R}_N\rbrace$.
According to \cite[Lemma 1]{CatSa10} the matrix $\bm{P}\left(\bm{I}_{Nm}-\mathcal{M}\mathcal{D}\right)$ is a stable matrix,
i.e.,  all its eigenvalues lie inside the unit circle.
Iterating (\ref{eqcit}) we obtain
\begin{align*}
\Vert E[\tilde{\underline{\bm{h}}}(n)]\Vert_{\ell_2} &\leq \lambda_{\max}\left(\bm{P}\left(\bm{I}_{Nm}-\mathcal{M}\mathcal{D}\right)\right)\Vert E[\tilde{\underline{\bm{h}}}(n-1)]\Vert_{\ell_2}  \\
&\leq \lambda_{\max}\left(\bm{P}\left(\bm{I}_{Nm}-\mathcal{M}\mathcal{D}\right)\right) ^{n-n_0}\Vert E[\tilde{\underline{\bm{h}}}(n_0)]\Vert_{\ell_2} 
\rightarrow 0, \ n\rightarrow\infty.
\end{align*}
The last relation completes our proof.
\section{Steady State Analysis}
Fix an arbitrary  node $k\in\N$, by employing the triangle inequality, it holds for the local error that
\begin{align}
\Vert \boldsymbol{\psi}_k(n)-\wo_* \Vert_{\ell_{2}} &\leq  \Vert \left(\boldsymbol{\psi}_k(n)-\wo_*\right)_{|\Ss_{k,n}} \Vert\nonumber\\&+\Vert \left(\boldsymbol{\psi}_k(n)-\wo_*\right)_{|\Ss_{k,n}^{c}} \Vert
\label{eq2}
\end{align}
Let us analyse each individual term of the right hand side of (\ref{eq2}). 
We obtain the following cases:
\begin{itemize}
\item $\Vert \wo_k(n-1)\Vert\leq D$:

Following
identical steps as in \cite[Thm. 3.8]{Fouc11} it holds that 
\begin{align}
\Vert \left(\boldsymbol{\psi}_k(n)-\wo_*\right)_{|\Ss_{k,n}^{c}}\|_{\ell_2}&\leq \sqrt{2}\Big[\delta_{3s,k}(\lambda,n)\|\boldsymbol{h}_k(n-1)\nonumber\\&-\wo_*\|_{\ell_2}+
\Vert\overline{\boldsymbol{\eta}}_k(n)\Vert_{\ell_2}\Big]\label{Eq:fouc_c}
\end{align}

\item $\Vert \wo_k(n-1)\Vert > D$.

Again, following similar steps as in \cite[Thm. 3.8]{Fouc11} we obtain that: 
 \begin{align}
&\Vert \left(\boldsymbol{\psi}_k(n)-\wo_*\right)_{|\Ss_{k,n}^{c}}\|_{\ell_2}\leq \nonumber\\&\sqrt{2}\left[\delta_{3s,k}(\lambda,n)\left\Vert \frac{\boldsymbol{h}_k(n-1)}{\Vert \boldsymbol{h}_k(n-1) \Vert}-\wo_*\right\Vert_{\ell_2}+
\Vert\overline{\boldsymbol{\eta}}_k(n)\Vert_{\ell_2}\right].
\label{Eq:fouc_c2}
\end{align}
Recall that  $\Vert \wo_k(n-1)\Vert >D \linebreak\geq 2 \Vert\wo_* \Vert_{\ell_2}+1$. Exploiting the triangle inequality, it holds that
\begin{equation*}
\left\Vert \frac{\boldsymbol{h}_k(n-1)}{\Vert \boldsymbol{h}_k(n-1) \Vert}-\wo_*\right\Vert_{\ell_2} \leq 1+ \Vert \wo_* \Vert_{\ell_2},
\end{equation*}
and
\begin{align*}
\|{\boldsymbol{h}_k(n-1)}&-\wo_*\|_{\ell_2} \geq \|{\boldsymbol{h}_k(n-1)}\|_{\ell_2}\nonumber\\ &- \Vert \wo_* \Vert_{\ell_2} \geq \Vert \wo_* \Vert_{\ell_2}+1. 
\end{align*}
The previous inequalities imply that 
\begin{equation*}
\left\Vert\frac{\boldsymbol{h}_k(n-1)}{\Vert \boldsymbol{h}_k(n-1) \Vert}-\wo_*\right\Vert_{\ell_2} \leq  \|{\boldsymbol{h}_k(n-1)}-\wo_*\|_{\ell_2},
\end{equation*}
and combining this with (\ref{Eq:fouc_c2}), we obtain that
\begin{align*}
&\Vert \left(\boldsymbol{\psi}_k(n)-\wo_*\right)_{|\Ss_{k,n}^{c}}\|_{\ell_2}\nonumber\\&\leq \sqrt{2}\Big[\delta_{3s,k}(\lambda,n)\|\boldsymbol{h}_k(n-1)-\wo_*\|_{\ell_2}\nonumber\\&+
\sqrt{1+\delta_{2s,k}(\lambda,n)}\Vert\overline{\boldsymbol{\eta}}_k(n)\Vert_{\ell_2}\Big].
\end{align*}
\end{itemize}
Hence, the inequality in (\ref{Eq:fouc_c2}) holds for both the proxy selection vectors.

\begin{remrk} The proxy noise disturbance $\overline{\boldsymbol{\eta}}_k(n)$ involves a double averaging (across neighbour and exponential average on each time instant) of two independent variables.
\end{remrk}


Next we analyse the first term of the right hand side of Eq. \eqref{eq2}. At each iteration, the Least--Mean-Square recursive relation (restricted at the local estimated support set $\Ss_{k,n}$)
\begin{align}
	\boldsymbol{\psi}_{k|\Ss_{k,n}}(n)&=\boldsymbol{h}_{k|\Ss_{k,n}}(n-1)\nonumber\\&+\mu_k\boldsymbol{a}_{k|\Ss_{k,n}}(n)\left[y_k(n)-\boldsymbol{a}^{T}_{k|\Ss_{k,n}}(n)\boldsymbol{h}_{k|\Ss_{k,n}}(n-1)\right]
\end{align}
updates the coefficient vector. Next we rewrite the restricted LMS iteration in term of the error vector $(\boldsymbol{\psi}_{k}(n)-\wo_*)_{k|\Ss_{k,n}}$ to obtain
\begin{align}
	&(\boldsymbol{\psi}_{k}(n)-\wo_*)_{k|\Ss_{k,n}}=\left(\boldsymbol{h}_{k}(n-1)-\wo_{*}\right)_{|\Ss_{k,n}} \nonumber\\&+\mu_k \boldsymbol{a}_{k|\Ss_{k,n}}(n)\left[y_k(n)-\boldsymbol{a}^{T}_{k|\Ss_{k,n}}(n)\boldsymbol{h}_{k|\Ss_{k,n}}(n-1)\right]\nonumber\\
	&=\left(\boldsymbol{h}_{k}(n-1)-\wo_{*}\right)_{|\Ss_{k,n}}+\mu_k \boldsymbol{a}_{k|\Ss_{k,n}}(n)\left[y_k(n)-\boldsymbol{a}^{T}_{k}(n)\wo_{*}\right]\nonumber\\
	&+\mu_k \boldsymbol{a}_{k|\Ss_{k,n}}(n)\left[\boldsymbol{a}^{T}_{k}(n)\wo_*-\boldsymbol{a}^{T}_{k|\Ss_{k,n}}(n)\boldsymbol{h}_{k|\Ss_{k,n}}(n-1)\right]\nonumber\\
	&=\left(\boldsymbol{I}_m-\mu_k\boldsymbol{a}_{k|\Ss_{k,n}}(n)\boldsymbol{a}^{T}_{k|\Ss_{k,n}}(n)\right)\left(\boldsymbol{h}_{k}(n-1)-\wo_{*}\right)_{|\Ss_{k,n}}\nonumber\\
	&+\mu_k \boldsymbol{a}_{k|\Ss_{k,n}}(n)e_{0,k}(n)+\mu_k \boldsymbol{a}_{k|\Ss_{k,n}}(n)\boldsymbol{a}^{T}_{k|\So\setminus\Ss_{k,n}}(n)\wo_{*|\So\setminus\Ss_{k,n}}\label{Eq:ss_init2}	
\end{align}
where $e_{0,k}(n):=y_{k}(n)-\boldsymbol{a}^T_k(n)\wo_*$ in the adaptive filtering literature is referred to as the estimation error of the optimum Wiener solution $\boldsymbol{h}_{*}$ to the normal equations
\begin{align}
	\boldsymbol{R}_k\boldsymbol{h}=\boldsymbol{p}_k
\end{align}
with $\boldsymbol{R}_k:=\mathbb{E}[\boldsymbol{a}_{k}(n)\boldsymbol{a}_{k}^{T}(n)]$ and $\boldsymbol{p}_k:=\mathbb{E}[y_{k}(n)\boldsymbol{a}_{k}(n)]$. The recursion of Eq. \eqref{Eq:ss_init2} can be written in terms of the weight--error--vectors
\begin{align}
	\epsilon_k^{(1)}(n):=\|\boldsymbol{\psi}_k(n)-\wo_*\|_{\ell_2}\\
	\epsilon_k^{(2)}(n):=\|\boldsymbol{h}_k(n)-\wo_*\|_{\ell_{2}}
\end{align}
Additionally we invoke the \textit{Direct--Averaging} method \cite{Hay96}, provided that the step size $\mu_k$ is small, where the regression correlation terms $\boldsymbol{a}_{k}(n)\boldsymbol{a}_{k}^{T}(n)$ in Eq. \eqref{Eq:ss_init2} are approximated with its average (the autocorrelation matrix $\boldsymbol{R}_k$). Hence, Eq. \eqref{Eq:ss_init2} becomes
\begin{align}
	\epsilon_{k|\Ss_{k,n}}^{(1)}(n)&\leq (1-\mu_k\lambda_{min})\epsilon_{k|\Ss_{k,n}}^{(2)}(n-1)+\mu_k\|\boldsymbol{a}_{k|\Ss_{k,n}}^{T}\|_{\ell_2}(n)|e_{o,k}(n)|\nonumber\\
	&+\mu_k \|\boldsymbol{a}_{k|\Ss_{k,n}}(n)\boldsymbol{a}^{T}_{k|\So\setminus\Ss_{k,n}}(n)\|_{\ell_2}\|\wo_{*|\So\setminus\Ss_{k,n}}\|_{\ell_2}\label{Eq:ss3}
\end{align}
and $\lambda_{min}$ is the minimum eigenvalue of $\boldsymbol{R}_{k}$ \cite[p. 168]{Hay96}, assuming that that it is non singular (\textit{i.e.} $\lambda_{min}>0$). Note that unlike the batch variants of the proposed algorithm (into their centralized form, \textit{i.e.} \cite{Fouc11,BluDa09}), the proposed algorithm is not applied to a fixed block of measurements. Therefore, the analysis of the proposed algorithm needs to take into account time dependencies at the support of $\boldsymbol{h}_k(n)$. The changes in the support of the local estimate across different $n$ are considered via the following expression: 
\begin{align}
	\epsilon_{k|\Ss_{k,n}}^{(1)}(n)\nonumber\\\nonumber &\hspace{-35pt}\leq (1-\mu_k\lambda_{min})\left\|\left(\boldsymbol{h}_k(n-1)-\wo_*\right)_{|\Ss_{k,n}}+\left(\boldsymbol{h}_k(n-1)-\wo_*\right)_{|\Ss_{k,n-1}}\right.\nonumber\\
	&\hspace{-35pt}\left.+\left(\wo_*-\boldsymbol{h}_k(n-1)\right)_{|\Ss_{k,n-1}}\right\|_{\ell_2}+\mu_k\|\boldsymbol{a}_{k|\Ss_{k,n}}^{T}(n)\|_{\ell_2}|e_{o,k}(n)|\nonumber\\
	&+\mu_k \|\boldsymbol{a}_{k|\Ss_{k,n}}(n)\boldsymbol{a}^{T}_{k|\So\setminus\Ss_{k,n}}(n)\|_{\ell_2}\|\wo_{*|\So\setminus\Ss_{k,n}}\|_{\ell_2}\label{Eq:ss4}\\
	\epsilon_{k|\Ss_{k,n}}^{(1)}(n)&\leq (1-\mu_k\lambda_{min})\left\{\epsilon_{k|\Ss_{k,n}}^{(2)}(n-1)+2\epsilon_{k|\Ss_{k,n-1}}^{(2)}(n-1)\right\}\nonumber\\
	&+\mu_k\|\boldsymbol{a}_{k|\Ss_{k,n}}^{T}(n)\|_{\ell_2}|e_{o,k}(n)|\nonumber\\&+\mu_k \|\boldsymbol{a}_{k|\Ss_{k,n}}(n)\boldsymbol{a}^{T}_{k|\So\setminus\Ss_{k,n}}(n)\|_{\ell_2}\|\wo_{*|\So\setminus\Ss_{k,n}}\|_{\ell_2}\label{Eq:ss5}
\end{align}
Now we want to express the fourth term of Eq. \eqref{Eq:ss5} in terms of $\epsilon_{k}^{(2)}(n-1)$. From Theorem 3.8 of \cite{Fouc11} and the direct--averaging method, one can write
\begin{align}
&\|\boldsymbol{a}_{k|\Ss_{k,n}}(n)\boldsymbol{a}^{T}_{k|\So\setminus\Ss_{k,n}}(n)\|_{\ell_2} \|\wo_{*|\So\setminus\Ss_{k,n}}\|\nonumber\\&\leq \|\boldsymbol{R}_{k}\|_{\ell_2}\|\left(\boldsymbol{\psi}_k(n)-\wo_*\right)_{\Ss^{c}_{k,n}}\|_{\ell_2}\nonumber= \lambda_{\max}\epsilon_{k|\Ss^{c}_{k,n}}^{(1)}(n)\nonumber\\
\leq & \sqrt{2}\lambda_{\max}\left[\delta_{3s,k}(\lambda,n)\epsilon_{k}^{(2)}(n-1)+
\sqrt{1+\delta_{2s,k}(\lambda,n)}\Vert\overline{\boldsymbol{\eta}}_k(n)\Vert_{\ell_2}\right]\label{Eq:fouc_c2a}
\end{align}

By combining Eqs. \eqref{Eq:ss5}, \eqref{Eq:fouc_c2a}, \eqref{Eq:fouc_c} in Eq. \eqref{eq2} we obtain the following recurrence 
\begin{align}
	\epsilon_{k}^{(1)}(n)&\leq (1-\mu_k\lambda_{min})\left\{\epsilon_{k|\Ss_{k,n}}^{(2)}(n-1)+2\epsilon_{k|\Ss_{k,n-1}}^{(2)}(n-1)\right\}\nonumber\\
	&+\sqrt{2}\delta_{3s,k}(\lambda,n)\left(1+\mu_k\lambda_{\max}\right)\epsilon_{k}^{(2)}(n-1)\nonumber\\
	&+\sqrt{2}\sqrt{1+\delta_{2s,k}(\lambda,n)}\left(1+\mu_k\lambda_{\max}\right)\Vert\overline{\boldsymbol{\eta}}_k(n)\Vert_{\ell_2}\nonumber\\
	&+\mu_k\|\boldsymbol{a}_{k|\Ss_{k,n}}^{T}(n)\|_{\ell_2}|e_{o,k}(n)|
\end{align}
Note that, 
the following two inequalities hold: $\epsilon_{k|\Ss_{k,n-1}}^{(2)}(n-1)\leq \epsilon_{k}^{(2)}(n-1)$ and $\epsilon_{k|\Ss_{k,n}}^{(2)}(n-1)\leq \epsilon_{k}^{(2)}(n-1)$ so that
\begin{align}
	\epsilon_{k}^{(1)}(n)&\leq \Big[3(1-\mu_k\lambda_{min})\nonumber\\&+\sqrt{2}\delta_{3s,k}(\lambda,n)\left(1+\mu_k\lambda_{\max}\right)\Big]\epsilon_{k}^{(2)}(n-1)\nonumber\\
	&+\sqrt{2}\sqrt{1+\delta_{2s,k}(\lambda,n)}\left(1+\mu_k\lambda_{\max}\right)	\Vert\overline{\boldsymbol{\eta}}_k(n)\Vert_{\ell_2}\nonumber\\
	&+\mu_k\|\boldsymbol{a}_{k|\Ss_{k,n}}^{T}(n)\|_{\ell_2}|e_{o,k}(n)|
\end{align}
Finally, for the whole network we have
\begin{align}
	&\underline{\epsilon}(n)\leq 2N\max_{k\in\mathcal{N}}\Big\{3(1-\mu_k\lambda_{min})+\sqrt{2}\delta_{3s,k}(\lambda,n)\nonumber\\&\left(1+\mu_k\lambda_{\max}\right)\Big\}\sum_{k=1}^N \epsilon_{k}^{(2)}(n-1)\nonumber\\
	&+2N\max_{k\in\mathcal{N}}\left\{\sqrt{2}\sqrt{1+\delta_{2s,k}(\lambda,n)}\left(1+\mu_k\lambda_{\max}\right)\right\}\sum_{k=1}^{N}\Vert\overline{\boldsymbol{\eta}}_k(n)\Vert_{\ell_2}\nonumber\\
	&+2N\max_{k\in\mathcal{N}}\left\{\mu_k\|\boldsymbol{a}_{k|\Ss_{k,n}}^{T}(n)\|_{\ell_2}\right\}\sum_{k=1}^{N}|e_{o,k}(n)|\label{Eq:final_trans}
\end{align}
We arrived at Eq. \eqref{Eq:final_trans} by recalling Remark 3. The second term on the right hand side of the Eq. \eqref{Eq:final_trans} reminds us the steady--state error of the HTP algorithm \cite{Fouc11} (mainly because we are using an exponentially weighted average proxy variant of \cite{Fouc11}), whereas the first and second term are introduced due to the usage of the diffusion LMS algorithm instead of the LS estimator.

 \bibliographystyle{IEEEbib}
\bibliography{ref4}

\begin{thebibliography}{10}

\bibitem{BrDoEl09}
A.~M. Bruckstein, D.~L. Donoho, and M.~Elad,
\newblock ``From sparse solutions of systems of equations to sparse modeling of
  signals and images,''
\newblock {\em SIAM review}, vol. 51, no. 1, pp. 34--81, 2009.

\bibitem{ThKoSl13}
S.~Theodoridis, Y.~Kopsinis, and K.~Slavakis,
\newblock ``Sparsity--aware learning and compressed sensing: An overview,''
\newblock {\em arXiv preprint arXiv:1211.5231}, 2012.

\bibitem{M10}
M.~Elad,
\newblock {\em Sparse and Redundant Representations},
\newblock Springer, 2010.

\bibitem{babadi2010sparls}
Behtash Babadi, Nicholas Kalouptsidis, and Vahid Tarokh,
\newblock ``Sparls: The sparse {RLS} algorithm,''
\newblock {\em IEEE Transactions on Signal Processing}, vol. 58, no. 8, pp.
  4013--4025, 2010.

\bibitem{CMG14}
S.J.~Godsill A.Y.~Carmi, L.~Mihaylova, Ed.,
\newblock {\em Compressed Sensing \& Sparse Filtering},
\newblock Springer, 2014.

\bibitem{Theo15}
S.~Theodoridis,
\newblock {\em Machine Learning: A Signal and Information Processing and
  Analysis Perspective},
\newblock Academic Press, 2015.

\bibitem{Lin13}
Jimmy Lin,
\newblock ``Mapreduce is good enough? {I}f all you have is a hammer, throw away
  everything that's not a nail!,''
\newblock {\em Big Data}, vol. 1, no. 1, pp. 28--37, 2013.

\bibitem{boyd11}
S.~Boyd, N.~Parikh, E.~Chu, B.~Peleato, and J.~Eckstein,
\newblock ``Distributed optimization and statistical learning via the
  alternating direction method of multipliers,''
\newblock {\em Foundations and Trends in Machine Learning}, vol. 3, no. 1, pp.
  1--122, 2011.

\bibitem{TW10}
J.A. Tropp and S.J. Wright,
\newblock ``Computational methods for sparse solution of linear inverse
  problems,''
\newblock {\em Proceedings of the IEEE}, vol. 98, no. 6, pp. 948 --958, 2010.

\bibitem{Tr04}
Joel~A Tropp,
\newblock ``Greed is good: Algorithmic results for sparse approximation,''
\newblock {\em IEEE Transactions on Information Theory}, vol. 50, no. 10, pp.
  2231--2242, 2004.

\bibitem{DTDS09}
David~L Donoho, Yaakov Tsaig, Iddo Drori, and J-L Starck,
\newblock ``Sparse solution of underdetermined systems of linear equations by
  stagewise orthogonal matching pursuit,''
\newblock {\em IEEE Transactions on Information Theory}, vol. 58, no. 2, pp.
  1094--1121, 2012.

\bibitem{NeedTro09}
D.~Needell and J.A. Tropp,
\newblock ``Cosamp: Iterative signal recovery from incomplete and inaccurate
  samples,''
\newblock {\em Applied and Computational Harmonic Analysis}, vol. 26, no. 3,
  pp. 301 -- 321, 2009.

\bibitem{DaMi09}
Wei Dai and Olgica Milenkovic,
\newblock ``Subspace pursuit for compressive sensing signal reconstruction,''
\newblock {\em IEEE Trans. on Information Theory}, vol. 55, no. 5, pp.
  2230--2249, 2009.

\bibitem{NV09}
D.~Needell and R.~Vershynin,
\newblock ``Uniform uncertainty principle and signal recovery via regularized
  orthogonal matching pursuit,''
\newblock {\em Found. Comput. Math}, vol. 9, no. 3, pp. 317--334, 2009.

\bibitem{HM11}
Honglin Huang and A.~Makur,
\newblock ``Backtracking-based matching pursuit method for sparse signal
  reconstruction,''
\newblock {\em IEEE Signal Processing Letters}, vol. 18, no. 7, pp. 391--394,
  2011.

\bibitem{PEE12}
T.~Peleg, Y.C. Eldar, and M.~Elad,
\newblock ``Exploiting statistical dependencies in sparse representations for
  signal recovery,''
\newblock {\em IEEE Transactions on Signal Processing}, vol. 60, no. 5, pp.
  2286--2303, 2012.

\bibitem{Tem11}
V.~Temlyakov,
\newblock {\em Greedy Approximation},
\newblock Cambridge University Press, 2011.

\bibitem{MaBaGi10}
G.~Mateos, J.A. Bazerque, and G.B. Giannakis,
\newblock ``Distributed sparse linear regression,''
\newblock {\em IEEE Trans. Signal Process.}, vol. 58, no. 10, pp. 5262--5276,
  2010.

\bibitem{MoXaAqPu10}
Jo{\~a}o~FC Mota, Jo{\~a}o~MF Xavier, Pedro~MQ Aguiar, and Markus P{\"u}schel,
\newblock ``Distributed basis pursuit,''
\newblock {\em IEEE Transactions on Signal Processing}, vol. 60, no. 4, pp.
  1942--1956, 2012.

\bibitem{PaElKe13}
Stacy Patterson, Yonina~C Eldar, and Idit Keidar,
\newblock ``Distributed sparse signal recovery for sensor networks,''
\newblock in {\em IEEE International Conference on Acoustics, Speech and Signal
  Processing (ICASSP)}. IEEE, 2013, pp. 4494--4498.

\bibitem{LoSa12}
P.~Di~Lorenzo and A.H. Sayed,
\newblock ``Sparse distributed learning based on diffusion adaptation,''
\newblock {\em IEEE Transactions on Signal Processing}, vol. 61, no. 6, pp.
  1419--1433, 2013.

\bibitem{ChoSlaKoT12}
S.~Chouvardas, K.~Slavakis, Y.~Kopsinis, and S.~Theodoridis,
\newblock ``A sparsity promoting adaptive algorithm for distributed learning,''
\newblock {\em IEEE Trans. Signal Process.}, vol. 60, no. 10, pp. 5412 --5425,
  2012.

\bibitem{liu2014distributed}
Zhaoting Liu, Ying Liu, and Chunguang Li,
\newblock ``Distributed sparse recursive least-squares over networks,''
\newblock {\em IEEE Transactions on Signal Processing}, vol. 62, pp.
  1386--1395, 2014.

\bibitem{CKMJXM02}
Chris Clifton, Murat Kantarcioglu, Jaideep Vaidya, Xiaodong Lin, and Michael~Y
  Zhu,
\newblock ``Tools for privacy preserving distributed data mining,''
\newblock {\em ACM SIGKDD Explorations Newsletter}, vol. 4, no. 2, pp. 28--34,
  2002.

\bibitem{sayedarxiv}
Ali~H Sayed,
\newblock ``Diffusion adaptation over networks,''
\newblock {\em arXiv preprint arXiv:1205.4220}, 2012.

\bibitem{BeTs99}
Dimitri~P Bertsekas and John~N Tsitsiklis,
\newblock {\em Parallel and distributed computation: Numerical Methods},
\newblock Athena-Scientific, second edition, 1999.

\bibitem{FoCaGi10}
Pedro~A Forero, Alfonso Cano, and Georgios~B Giannakis,
\newblock ``Consensus-based distributed support vector machines,''
\newblock {\em The Journal of Machine Learning Research}, vol. 99, pp.
  1663--1707, 2010.

\bibitem{Fouc11}
S.~Foucart,
\newblock ``Hard thresholding pursuit: An algorithm for compressive sensing,''
\newblock {\em SIAM Journal on Numerical Analysis}, vol. 49, no. 6, pp.
  2543--2563, 2011.

\bibitem{LoSa08}
C.G. Lopes and A.H. Sayed,
\newblock ``Diffusion least-mean squares over adaptive networks: Formulation
  and performance analysis,''
\newblock {\em IEEE Trans. Signal Process.}, vol. 56, no. 7, pp. 3122--3136,
  2008.

\bibitem{CatSa10}
F.S. Cattivelli and A.H. Sayed,
\newblock ``Diffusion {LMS} strategies for distributed estimation,''
\newblock {\em IEEE Trans. Signal Process.}, vol. 58, no. 3, pp. 1035 --1048,
  2010.

\bibitem{ChoSlTh12}
S.~Chouvardas, K.~Slavakis, and S.~Theodoridis,
\newblock ``Adaptive robust distributed learning in diffusion sensor
  networks,''
\newblock {\em IEEE Trans. Signal Process.g}, vol. 59, no. 10, pp. 4692--4707,
  2011.

\bibitem{ScMaGi09}
I.D. Schizas, G.~Mateos, and G.B. Giannakis,
\newblock ``Distributed {LMS} for consensus-based in-network adaptive
  processing,''
\newblock {\em IEEE Trans. Signal Process.}, vol. 57, no. 6, pp. 2365 --2382,
  2009.

\bibitem{TuSa12}
S.Y. Tu and A.H. Sayed,
\newblock ``Diffusion networks outperform consensus networks,''
\newblock in {\em IEEE Workshop of SSP}, 2012, pp. 313--316.

\bibitem{Say08}
A.H. Sayed,
\newblock {\em Adaptive Filters},
\newblock John Wiley and Sons, 2008.

\bibitem{MiBaKaTa10}
G.~Mileounis, B.~Babadi, N.~Kalouptsidis, and V.~Tarokh,
\newblock ``An adaptive greedy algorithm with application to nonlinear
  communications,''
\newblock {\em IEEE Trans. Signal Process.}, vol. 58, no. 6, pp. 2998 --3007,
  2010.

\bibitem{XiBo04}
L.~Xiao and S.~Boyd,
\newblock ``Fast linear iterations for distributed averaging,''
\newblock {\em Systems \& Control Letters}, vol. 53, no. 1, pp. 65--78, 2004.

\bibitem{CaYaMu09}
Renato~LG Cavalcante, Isao Yamada, and Bernard Mulgrew,
\newblock ``An adaptive projected subgradient approach to learning in diffusion
  networks,''
\newblock {\em Signal Processing, IEEE Transactions on}, vol. 57, no. 7, pp.
  2762--2774, 2009.

\bibitem{PaPi02}
Athanasios Papoulis and S~Unnikrishna Pillai,
\newblock {\em Probability, random variables, and stochastic processes},
\newblock Tata McGraw-Hill Education, 2002.

\bibitem{Hay96}
S.~Haykin,
\newblock {\em Adaptive Filter Theory},
\newblock Prentice Hall, 1996.

\bibitem{LoBaSa13}
P.~Di~Lorenzo, S.~Barbarossa, and A.H. Sayed,
\newblock ``Distributed spectrum estimation for small cell networks based on
  sparse diffusion adaptation,''
\newblock {\em IEEE Signal Processing Letters}, vol. 20, no. 12, pp.
  1261--1265, Dec 2013.

\bibitem{BluDa09}
Thomas Blumensath and Mike~E. Davies,
\newblock ``Iterative hard thresholding for compressed sensing,''
\newblock {\em Applied and Computational Harmonic Analysis}, vol. 27, no. 3,
  pp. 265 -- 274, 2009.

\end{thebibliography}
\end{document}